\documentstyle[epsf,aas2pp4]{article}

\newcommand{\etal}{{et al.}\hspace{1mm}}

\begin{document}


\title{Resolving the Ly$\alpha$ Forest} 

\author{Greg L. Bryan\altaffilmark{1}}
\affil{Princeton University Observatory, Princeton, NJ 08544}

\author{Marie Machacek}
\affil{Department of Physics, Northeastern University, Boston, MA 02115}

\author{Peter Anninos}
\affil{Laboratory for Computational Astrophysics, National Center for
Supercomputing Applications, \\ 405 Mathews Ave., Urbana, IL 61801}

\and

\author{Michael L. Norman\altaffilmark{2}}
\affil{Astronomy Department, University of Illinois at
Urbana-Champaign, Urbana, IL 61801}

\altaffiltext{1}{Lyman Spitzer Jr. Fellow}
\altaffiltext{2}{also at National Center for Supercomputing
Applications, Urbana IL 61801}


\begin{abstract}

In this paper we critically examine predictions of the Ly$\alpha$
forest within the standard cold dark matter (SCDM) model, paying
particular attention to the low end of the column-density
distribution.  We show in particular that the width of these lines,
typically measured by the $b$-parameter of a Voigt profile, is
sensitive to spatial resolution in numerical simulations and has
previously been overestimated.  The new result, which predicts a
distribution with a median $b$ of around 20-22 km/s at $z=3$, is
substantially below that observed.
We examine a number of possible causes of this discrepancy and argue
that it is unlikely to be rectified by an increase in the thermal
broadening of the absorbing gas, but is instead telling us something
about the distribution of matter on these scales.  Although the median
differs, the shape of the $b$-parameter distribution agrees quite well
with that observed, and the high-end tail is naturally produced by the
filamentary nature of gravitational collapse in these models.  In
particular, we demonstrate that lines of sight which obliquely
intersect a filament or sheet tend to produce absorption lines with
larger $b$ parameters.  We also examine the physical nature of the gas
which is responsible for the forest, showing that for lines with
neutral column densities below $N_{HI} \sim 10^{14}$ cm$^{-2}$ (for
this model at $z=3$), the peculiar infall velocity is actually slower
than the Hubble flow, while larger lines have, on average, turned
around and are collapsing.

\end{abstract}


\twocolumn


\section{Introduction}

A physical picture of the Ly$\alpha$ forest in Cold Dark Matter (CDM)
dominated cosmologies has recently emerged from numerical simulations
(e.g. \cite{cen94}; \cite{zha95}; \cite{her96}), and other
approximation techniques (\cite{bi93}; \cite{bi97}; \cite{hui97};
\cite{gnehui97}).  In this context, the absorbers that give rise to
low column density lines ($N_{HI} < 10^{15}$ cm$^{-2}$) at $z \sim 3$ are
large, unvirialized objects with sizes of $\sim$ 100 kpc, and low
densities, comparable to the cosmic mean (\cite{zha97b}).  The width
of the lines, as measured by the $b$ parameter of a Voigt
profile is set not only by thermal broadening or peculiar
velocities, but also by the Hubble expansion across their width
(\cite{wei97}).  In such a situation, there is a monotonic
relationship between baryonic density and optical depth which can be
exploited to investigate the density distribution along the quasar
lines of sight (e.g. \cite{cro97}).

This explanation of the forest (for previous ideas along this
direction, see also \cite{bon88}, \cite{mcg90}, \cite{bi93}, and
\cite{mei94}), arising from gravitationally amplified primordial
fluctuations, allows us to test various models of cosmological
structure formation.  This can be done, for example, by comparing the
distribution of column densities (\cite{gne97}; \cite{bon97};
\cite{mac98}).  These studies have found that the overall
normalization of the distribution depends approximately on the
parameter $\Omega_b^2 h^3/\Gamma$ (there is also some dependence on
gas temperature).  Here $\Gamma$ is the HI ionization rate, $h$ is the
Hubble constant in units of 100 km/s/Mpc, and $\Omega_b$ is the ratio
of the baryon density to the critical density required to close the
universe.  For a given model, this parameter is often set by requiring
that the mean optical depth match that observed (\cite{pre93};
\cite{zuo93}; \cite{rau97}).  The distribution of column densities is
close to a power law and at least approximate
agreement seems
to be found for a number of popular cosmological models.  Another
diagnostic is the distribution of $b$ parameters found by fitting
Voigt profiles to the spectra.  This has been suggested as a probe of
the reionization history (\cite{hae97}).


In this paper, we undertake a systematic evaluation of numerical
uncertainties in simulations of the Ly$\alpha$ forest, and in doing
so, explore the physical structure of the lines.  We will show that
the line structure can be rather simply understood, but does require
relatively high spatial resolution to model accurately, particularly
for the fluctuations at or below the cosmic mean which give rise
to the low end of the column density distribution.  We cannot
address either Lyman-limit or damped Ly$\alpha$ systems as they
require more spatial resolution and physical processes than these
simulations provide.

The paper is structured in the following way: in
section~\ref{sec:resolution_intro}, we examine the effect of
resolution on various statistical measures of the forest and the gas
that gives rise to it, looking first at distributions of density and
temperature (section \ref{sec:physical}), and then at the properties
of the lines fit to simulated spectra, in section~\ref{sec:lines}.  We
analyze the physical nature of the lines in the next two sections
(\ref{sec:nature} and \ref{sec:jeans}), other non-parametric measures
of the spectra (section~\ref{sec:non_param}), and the effect of our
limited computational volume (section~\ref{sec:power}).  Then we
discuss these results and the corresponding observations in
section~\ref{sec:conclusions}.

While this paper was in the final stages of preparation, a preprint
(\cite{the98}) was circulated which examined some of the same issues,
although with a different numerical method.  Where there is overlap,
our results are in agreement.  In particular, they also find that the
b-parameter distribution requires very high resolution.


\section{The effect of spatial resolution}
\label{sec:resolution_intro}


Most of the results for this section come from a set of three
simulations in which we modeled the same region of space, but each time
increased the resolution by a factor of two.  In generating the
initial conditions, we keep constant not just the amplitudes but also
the phases of the initial conditions (although, of course, higher
resolution simulations contain more high-frequency power).


\subsection{The simulations}

The model is (nearly) the standard Cold Dark Matter (SCDM) model, with
total and baryon density parameters $\Omega_0=1$ and $\Omega_b=0.06$,
respectively.  The Hubble constant is $h=0.5$ and the matter
fluctuation spectrum was normalized to $\sigma_8=0.7$, which measures
the {\it rms} density fluctuations in a tophat sphere of 8$h^{-1}$
Mpc.  This normalization is considerably lower than the COBE measured
CMB fluctuations would imply, however it is more in line with
(although marginally higher than) the value indicated by the number
density of large galaxy clusters (\cite{via96}; \cite{whi93}).  We
choose this model because it is a standard which we have simulated
before (\cite{zha97a}; hereafter known as ZANM97).  The power spectrum
is realistic enough to produce results which are characteristic of
these kinds of hierarchical models, although certainly other
parameters can be found which provide a better match to other
available observations.

Our box size is 1.2$h^{-1}$ Mpc, too small to include much of the
large-scale power important to accurately determine the true
distribution of column densities and $b$ parameters, as we will show
below (see also \cite{wad97}, \cite{bon97}).  We adopt such a small
volume in order to obtain the highest possible resolution for the
lines we do resolve.  In other words, we sacrifice the correct
distribution of line widths and sizes under the assumption that the
internal structure of a given line is not affected by the loss of
large-wavelength power.  We will show some supporting evidence for
this with simulations of larger regions.  Three different resolutions
are computed, ranging from $32^3$ to $128^3$ cells, or cell sizes from
$37.5h^{-1}$ kpc to $9.375h^{-1}$ kpc.  

In order to set this work within the context of other simulations, we
show the cell size of some other representative Eulerian Ly$\alpha$
simulations in Table~\ref{table:simulations}, along with all the runs
used in this paper.  The first three lines are the resolution study
runs that will be discussed in the next five sections, while the next
five will be used to examine the effect of large-scale power
(section~\ref{sec:power}).  The gravitational softening length of
these simulations is typically 1.5--2 cell widths.  Also shown is the
Hubble flow across a cell at $z=3$, as well as the size of the box.
Of the three simulations by other authors cited in the table, only one
(\cite{zha97b}) is of the same cosmological model used in this paper.
The other two examined a cosmological-constant dominated cosmology
(\cite{mcor}; \cite{ost95}).


{
\begin{deluxetable}{ccccc}
\footnotesize
\tablecaption{Selected Eulerian Ly$\alpha$ forest simulations}
\tablewidth{0pt}
\tablehead{
\colhead{reference} & 
\colhead{$\Delta x$ ($h^{-1}$kpc)} & 
\colhead{$H(z) \Delta x/(1+z)$ at $z=3$ (km/s)} & 
\colhead{$L_{\hbox{box}}$ ($h^{-1}$ Mpc)}
}
\startdata
$32^3$ (this work) & 37.5 & 7.5 & 1.2 \nl
$64^3$ (this work) & 18.9 & 3.8 & 1.2 \nl
$128^3$ (this work) & 9.4 & 1.9 & 1.2 \nl
\tablevspace{0.15cm}
L1.2 (this work) & 37.5 & 7.5 & 1.2 \nl
L2.4 (this work) & 37.5 & 7.5 & 2.4 \nl
L4.8 (this work) & 37.5 & 7.5 & 4.8 \nl
L9.6 (this work) & 37.5 & 7.5 & 9.6 \nl
L4.8HR (this work) & 18.9 & 3.8 & 4.8 \nl
\tablevspace{0.15cm}
\cite{zha97b} top (sub) grid & 37.5 (9.4) & 7.5 (1.9) & 4.8 \nl 
\cite{mcor} L10 & 34.7 & 4.2 & 10 \nl
\cite{mcor} L3  & 10.4 & 1.3 & 3 \nl
\enddata
\label{table:simulations}
\end{deluxetable}
}


We do not include the smoothed-particle hydrodynamics (SPH)
simulations since their adaptive smoothing length makes a direct
comparison difficult.  However we note that the smoothing length of
typical SPH simulations scales approximately as the interparticle
spacing, which, for Hernquist \etal (1996)\markcite{her96}, is 174
$(\rho_b/\left<\rho_b\right>)^{-1/3} h^{-1}$ kpc, where
$\left<\rho_b\right>$ is the mean baryonic density.  For the mean
density at $z=3$, the Hubble flow across this distance is 34.8 km/s.
The gravitational smoothing lengths of SPH simulations are typically
much smaller (10 $h^{-1}$kpc for \cite{her96}).

The simulation technique uses a particle-mesh algorithm to follow the
dark matter and the piecewise parabolic method (PPM) to simulate the
gas dynamics (described in detail in \cite{PPM}).  Since
non-equilibrium effects can be important, six species are followed
(HI, HII, HeI, HeII, HeIII and the electron density) with a
sub-stepped backward finite-difference technique (\cite{ann97}).  We
assume a spatially-constant radiation field computed from the observed
QSO distribution (\cite{haa96} with an intrinsic spectral index of
$\alpha = 1.5$), which reionizes the universe around $z \sim 6$ and
peaks at $z \sim 2$.

As part of the analysis procedure, we generate spectra along random
lines of sight through the volume, including the effects of peculiar
velocity and thermal broadening of the gas (ZANM97).  The
spectra 
are fit by Gaussians (at these column densities, a Voigt-profile is
indistinguishable from a Gaussian in optical depth) to obtain column
densities and Doppler widths for each line.  To match observational
samples, a minimum optical depth at line center ($\tau_{HI} = 0.05$)
is required.  The procedure is described in more detail elsewhere
(ZANM97), but we go over it briefly here.  We neglect a number of
observational difficulties; specifically we do not include noise or
continuum-fitting.  First, maxima in the optical depth distribution
are identified as line centers.  Then, Gaussian profiles are fit,
using a non-linear minimization, to the part of the spectrum which is
above $\tau_{HI} = 0.05$ and between neighbouring minima.  The spectra
have a resolution of 1.2 km/s regardless of the spatial resolution of
the simulation, a value which is smaller than current observations.
We use this technique since we are not interested in a detailed
comparison to raw observations, but are more interested in exploring
the physical nature of the lines.  In other words, we assume that
observers have done a good job of correcting their samples for
observational biases.  Still, we will show (by direct comparison) that
the result of not including these observational effects is relatively
slight, but only as long as we restrict our comparison to high quality
observational samples.

\subsection{Physical properties}
\label{sec:physical}

\begin{figure}
\epsfxsize=3.5in
\centerline{\epsfbox{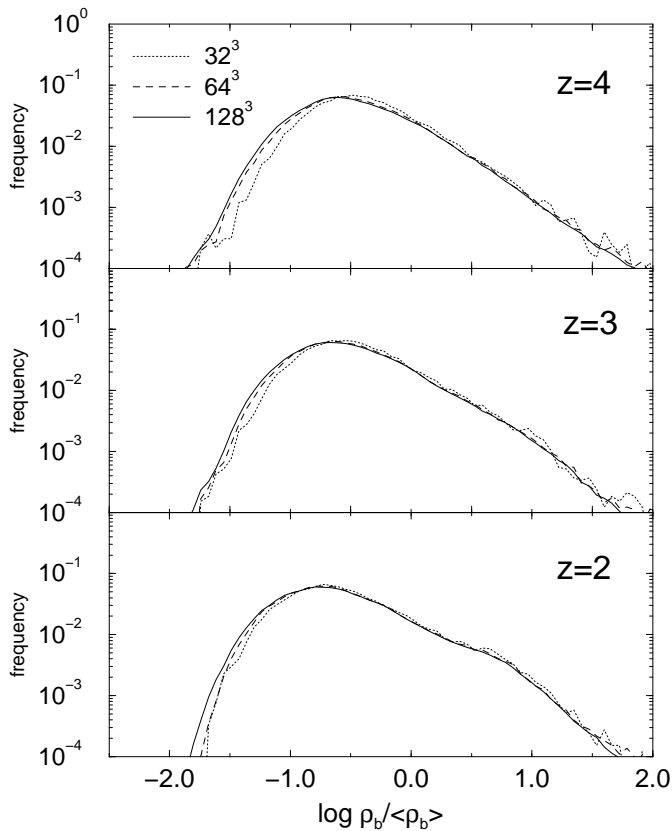}}
\caption{ 
The distribution of the volume-weighted baryon overdensity at
three redshifts ($z=2,3$, and 4) for three simulations with different
resolutions.}
\label{fig:dens_log_hist}
\end{figure}

\begin{figure*}
\epsfxsize=3in
\centerline{\epsfbox{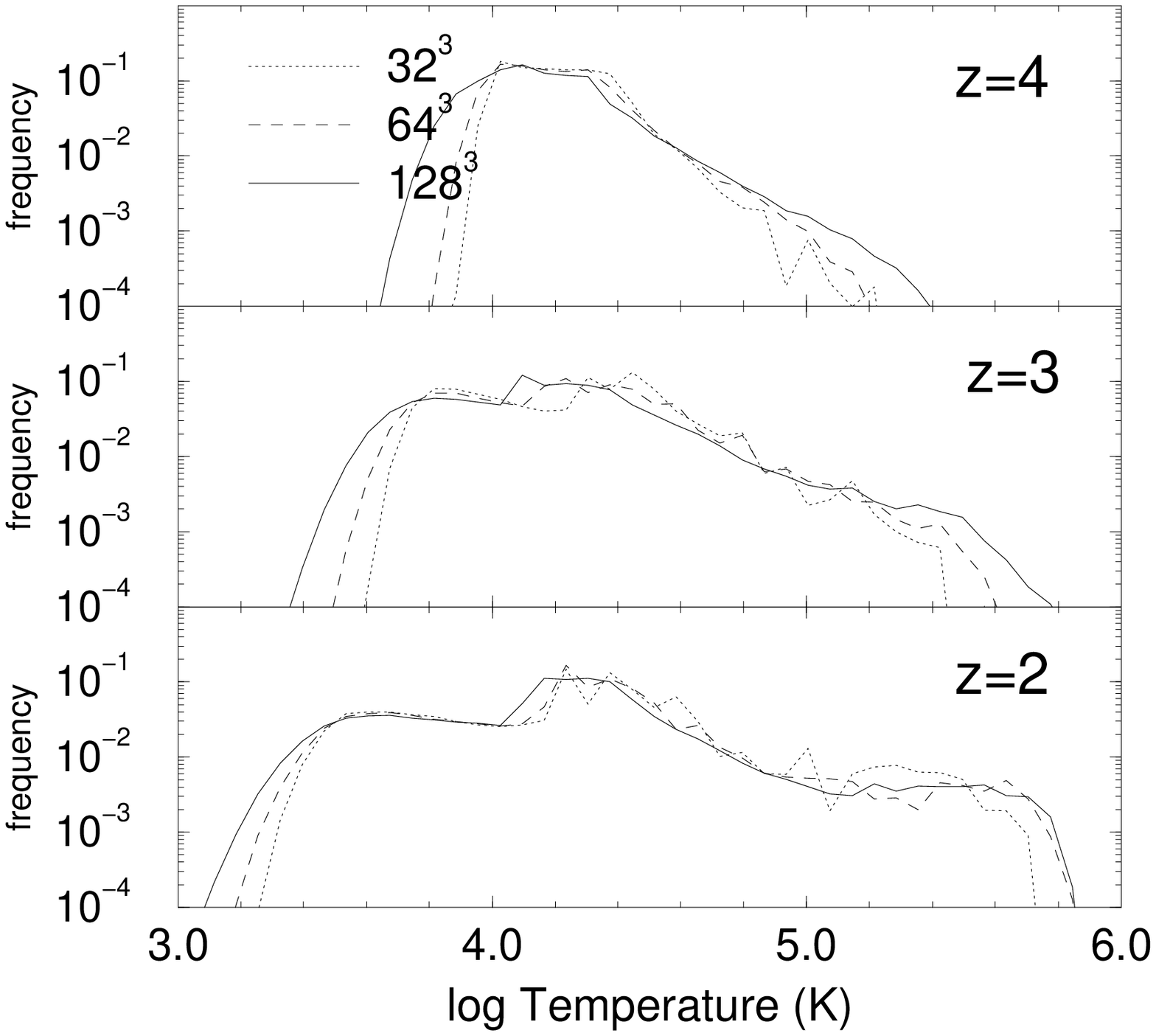}
\hspace{0.5cm}\epsfxsize=3in \epsfbox{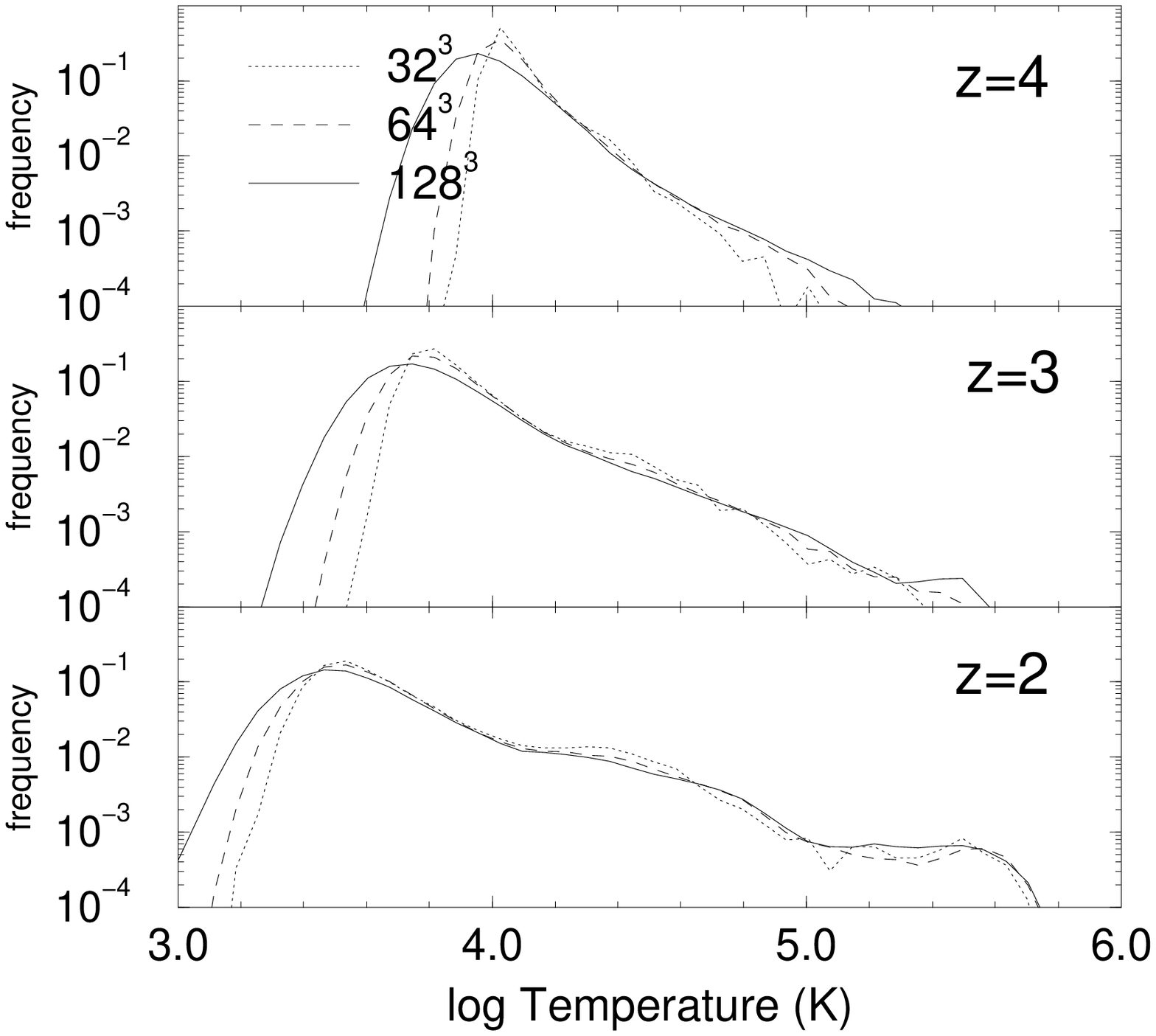}}
\caption{
The distribution of density-weighted temperature (left panels) and
volume-weighted temperature (right panels) at three redshifts
each for three resolutions.
}
\label{fig:temp_weight_log_hist}
\end{figure*}

In Figures~\ref{fig:dens_log_hist} and \ref{fig:temp_weight_log_hist},
we show the baryonic density and temperature distributions for the
three simulations at three redshifts.  The density distributions
follow the usual approximately log-normal curve (although note the
high-density tail), with the majority of the volume below the mean
density.  This pattern continues at lower redshifts as the voids
deepen and the filaments and peaks grow.  The difference between
resolutions is relatively small, although there is a slight tendency
for the higher resolution simulations to have more low-density cells,
due both to their additional small-scale power, and to the
improved resolution of the small-scale features already present.

The temperature distribution can be described as three features (seen
most clearly in the density-weighted distribution).  The central one
is a peak slightly above $10^4$ K, representing dense gas which has
collapsed, shock heated and then cooled, generally by line-emission,
to this temperature (below which emission processes for our assumed
metal-free gas are inefficient).  This virialized material is quite
dense but occupies a small volume as can be seen by
contrasting the volume and density weighted distributions.  This
means that, although they may be quite prominent in emission features
(and indeed would possibly undergo star formation if such processes
were included), most of the absorption lines, by number, come from
low-density regions that occupy the majority of the volume.

The second feature shows up in these plots as a low-temperature
shoulder which moves even further to the left at lower redshift.  This
is low-density gas which was originally heated to $\sim$ 15,000 K by
reionization occurring between $z=7$ and 6.  It then adiabatically
cools with the universal expansion and, because of the low HI fraction
and low recombination rate, ionization heating is insufficient after
$z\sim 5$ to keep this gas from cooling below $10^4$~K.  Most of it is
below the mean baryonic density.

The third feature is a long, power-law profile of shock-heated
material above $10^5$ K that is too low-density to cool efficiently.
It comes from the halos of clumps, filaments and, in some cases,
sheets.  The amount of such gas increases with time as longer
waves go non-linear, although it is clear that the volume
occupied by such gas is extremely small, and will not contribute
significantly to the bulk of the Ly$\alpha$ forest.  As we will
demonstrate in section~\ref{sec:power}, the sharp cutoff at $T \sim
10^5$--$10^6$ K is due to the artificial lack of large-scale power.

The effect of resolution is again quite small.  There are some
differences in the high-temperature end, particularly when looking at
the mass-weighted statistic, which is due to the improved resolution
around small collapsed objects.  More interestingly, from the
standpoint of the forest, higher resolution systematically produces
more low-temperature gas.  Such a trend is expected from the change in
the density distribution described above: lower density regions are
cooler due to enhanced adiabatic expansion and reduced
photo-ionization heating.

\begin{figure}
\epsfxsize=3in
\centerline{\epsfbox{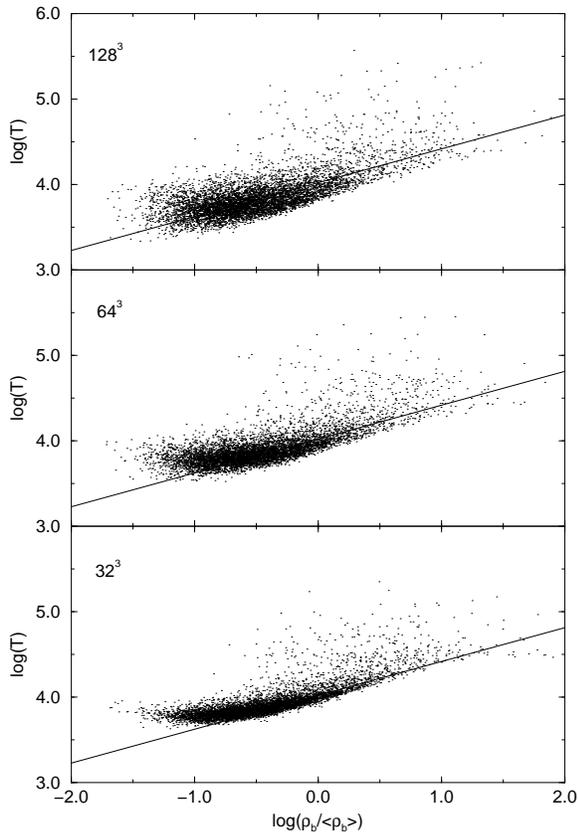}}
\caption{ 
The volume-weighted joint baryon density-temperature
distribution for three resolutions at $z=3$.  Each point is
one cell, subsampled to 8192 points for each panel.  The solid
line is the predicted equation of state as discussed in the text.
}
\label{fig:hist2d}
\end{figure}

A better understanding of the effect of resolution can be gained by
examining the joint density-temperature relation, shown in
Figure~\ref{fig:hist2d}.  The correlation between density and
temperature is due to the fact that the thermal history, primarily
photo-ionization heating and adiabatic cooling, depends only on
density for fluid elements that are not shocked.  At higher densities,
shocks play more of a role and the correlation is less well-defined.
We also show --- as a solid line --- the analytic equation of state as
predicted by Hui \& Gnedin (1997), for sudden reionization at $z=6$:
\begin{equation}
T = T_0 (\rho/\bar{\rho})^{\gamma -1}.
\end{equation}
where $T_0 = 10500$ and $\gamma = 1.4$.
As the resolution is improved, the scatter
around this relation increases (we will discuss the reason for this in
section~\ref{sec:jeans}).  This is of interest as much semi-analytic
work on the forest depends on this relation; however, it should be
kept in mind that this is a cell-by-cell plot, and an average over the
region that is actually observed as an absorption line will tend to
reduce the scatter.

\subsection{Line properties}
\label{sec:lines}

\begin{figure}
\epsfxsize=3.5in
\centerline{\epsfbox{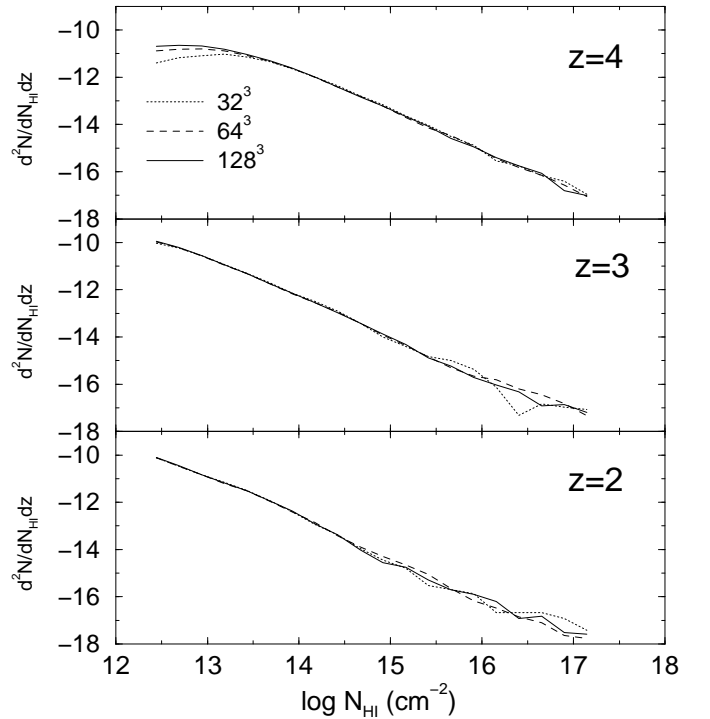}}
\caption{
The distribution of neutral hydrogen column densities for three
redshifts with three different resolutions, ranging from $32^3$ cells
(dotted line) in a $1.2h^{-1}$
Mpc box, to $128^3$ cells (sold line).
}
\label{fig:NH_dist}
\end{figure}

We have seen that there are some, albeit slight, changes to the gross
distribution functions of the physical variables in the simulations.
We ask here what change the resolution has on the observables of the
Ly$\alpha$ forest.  In Figure~\ref{fig:NH_dist}, we show the
distribution of column densities of neutral hydrogen for our three
different resolutions.  Although there are fluctuations at the higher
column density end due to shot noise from the small number of
absorbers, the only systematic differences are at high redshift, for
the lowest column densities, where observations are also uncertain.
This arises because this redshift and column density range probes
the smallest fluctuations present in the simulation (those which are
around the Jeans length, a fact which will be discussed in more
detail below).  As the cell width goes down, these fluctuations become
better resolved resulting in an increase in the number of these
low column-density lines.

Otherwise, there is remarkable agreement, demonstrating the forgiving
nature of the column density distribution.  Roughly speaking, this is
due to the fact that the column density of a given line is basically
an integrated quantity and is not sensitive to the detailed shape of
the gas profile.  This also comes into play at the high column density
end which comes from the dense, cooling knots (see also \cite{zha97b}).
At high resolutions these cores become smaller and denser but,
remarkably, this does not appear to substantially affect the column
density distribution because their cross-sections also decrease.  
Lines with column densities of around $10^{16.5}$ and higher suffer
from significant self-shielding, a process we have not included in the
simulations, so we do not go beyond this point.

\begin{figure}
\epsfxsize=3.5in
\centerline{\epsfbox{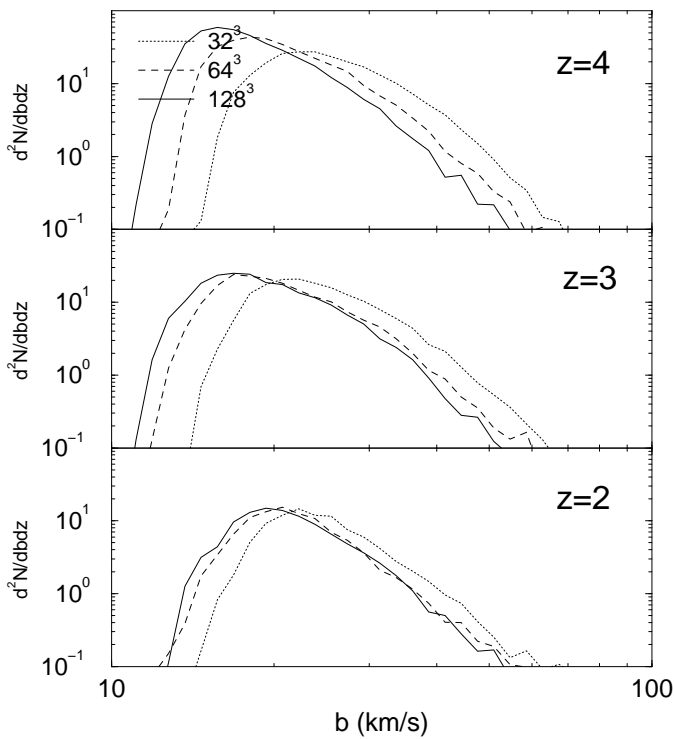}}
\caption{
The distribution of Doppler $b$ parameters (for lines with column
densities $10^{13} < N_{HI} < 10^{14}$ cm$^{-2}$) for the same three
redshifts and resolutions as in figure~\protect\ref{fig:NH_dist}.  Higher
resolution simulations produce distributions shifted to smaller $b$.
}
\label{fig:b_dist}
\end{figure}

The second quantity typically used to parameterize a Voigt profile is
the Doppler $b$ parameter which measures the width of the line.
Unlike the column density this parameter is sensitive to resolution,
as Figure~\ref{fig:b_dist} demonstrates.  We restrict ourselves to
column densities greater than $10^{13}$ cm$^{-2}$ and less than
$10^{14}$ cm$^{-2}$, since this range is well-constrained
observationally and does not suffer so much from selection effects due
to the minimum optical depth cut mentioned earlier (the behaviour of
lines with column densities outside of this range is quite similar).

The shape of the distribution may appear unfamiliar to those used to
seeing this quantity plotted linearly, but the profile is much like
that seen observationally.  The mean, in contrast, particularly for
the highest resolution case, is much smaller than observed, which is
around 30 km/s (\cite{hu95}; \cite{kir97}; \cite{kim97}).  This figure
is one of the main results of this paper, since it demonstrates that
(1) previous simulations have over-estimated the line widths, and (2)
this model, which has previously been shown to predict results (at
least for the Ly$\alpha$ forest) in agreement with observations, is
now discrepant.  Since it is of some significance, we will spend much
of the rest of the paper examining this point more closely.

\begin{figure}
\epsfxsize=3in
\centerline{\epsfbox{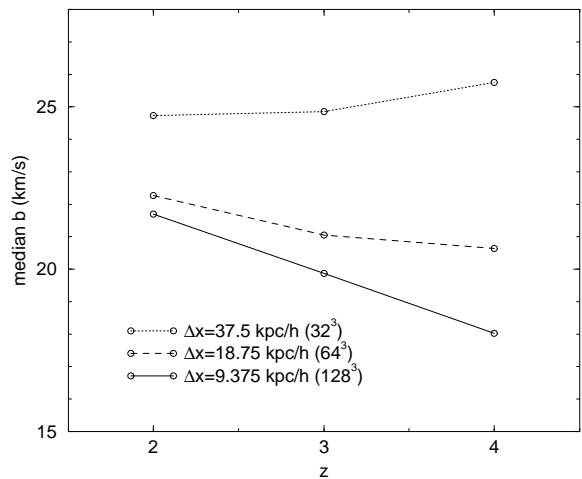}}
\caption{ 
The median of the Doppler $b$ distributions shown in
Figure~\protect\ref{fig:b_dist} as a function of redshift and
resolution (for lines in the $N_{HI} = 10^{13}$--$10^{14}$ cm$^{-2}$
range).  Higher resolution simulations produce thinner lines.  
}
\label{fig:b_median}
\end{figure}

The entire distribution tends to shift to the left as the resolution
improves, which means that each line decreases by a constant fraction
of its width.  Since the shape of the distribution is largely
invariant (we will explore this point in more detail later), we can
characterize the shift by the change in the median of the
distribution.  To show the trend more clearly,
Figure~\ref{fig:b_median} plots the median $b$ against resolution.
The drop at $z=3$ is about 5 km/s going from a comoving cell size of
37.5 $h^{-1}$ kpc to our highest resolution at 9.375 $h^{-1}$ kpc.
The effect is clearly largest at high redshift, and shows signs of
convergence by $z=2$.  We also note that for the highest resolution
simulations, there is a clear trend of increasing $b$ parameter with
decreasing redshift, as observations also seem to indicate
(\cite{kim97}), although of course with much larger median $b$'s.

\begin{figure}
\epsfxsize=3in
\centerline{\epsfbox{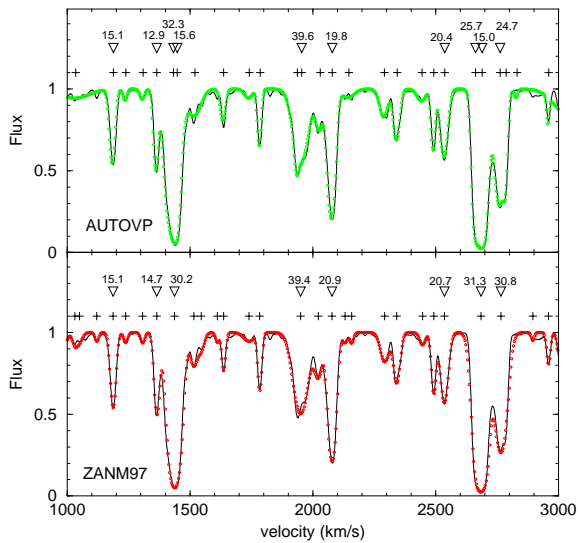}}
\caption{ 
A comparison of the same simulated spectrum fit with two different
techniques, one without noise (\protect\cite{zha97a}, denoted ZANM97
in the figure) and one with noise (AUTOVP, kindly provided by Romeel
Dav\'e).  The spectrum is from the $128^3$ simulation at $z=3$.  The
simulated spectrum is shown with a solid lines, while the fits are
circles.  Triangles
indicate fit lines with column densities greater than
$10^{13}$ cm$^{-2}$, and the associated numbers are the best fit $b$
parameters from each method.  Notice that these larger lines are well
reproduced and fit by both methods.
Plus symbols give the locations of lines with all column densities
found by the respective methods.
}
\label{fig:spect}
\end{figure}

An important question that needs to be addressed is whether such low
values are due to our rather idealized Voigt-profile fitting
procedure.  We test this possibility with the automated Voigt-profile
fitting package AUTOVP (\cite{dav97}), kindly provided to us by Romeel
Dav\'e.  The spectra are smoothed to approximate the resolution of
HIRES, and noise is added, both a readout component and a Gaussian
photon noise with a signal-to-noise ratio of 60.  An example of a
relatively small section of the spectrum at $z=3$ is shown in
Figure~\ref{fig:spect}.  Also shown, for lines greater than $10^{13}$ 
cm$^{-2}$, are the resulting $b$ parameters.  There is some indication
that AUTOVP adds more lines with somewhat lower $b$ values, which is
opposite to the bias required if our fitting algorithm is responsible
for the low median $b$ of the predicted distribution.  However, only
a few lines are shown, and a more statistical approach is required.

\begin{figure}
\epsfxsize=3in
\centerline{\epsfbox{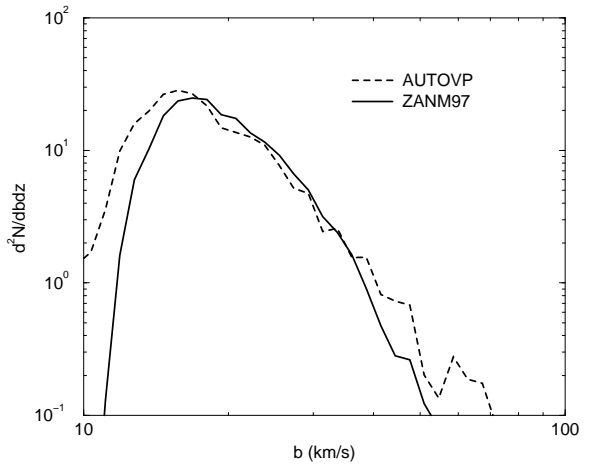}}
\caption{ 
A comparison of the Doppler $b$ distribution resulting from
fitting with two different techniques, one without noise
(ZANM97) and one with
noise (AUTOVP).  The spectra are from the $128^3$ simulation at $z=3$.
}
\label{fig:b_autovp}
\end{figure}

We fit one hundred
of our $\Delta z = 0.1$ spectra for a relatively difficult case, the
$128^3$ simulation at $z=3$, which exhibits many narrow lines.  The
result, shown in Figure~\ref{fig:b_autovp}, demonstrates that while
there are some differences --- which would be greater for lower
signal-to-noise spectra --- the basic outcome is unchanged.  Indeed,
this is unsurprising: the presence of metal lines in observed spectra
proves that there is no intrinsic difficulty in fitting thin lines.
The median $b$ parameters of the two distributions differ by only 1.7
km/s (19.9 km/s for the method of ZANM97 compared to 18.2 km/s for
AUTOVP).  This difference reflects the fact that Voigt-profile fitting
is not a highly robust way to fit these lines, but we argue it is
sufficient to differentiate between the spectra predicted here and
those observed.

Another possibility that we can examine is the effect of changing the
photo-ionization rate and hence the average flux decrement, which is
not all that well determined observationally: ranging from 0.22
(\cite{zuo93}) to 0.36 (\cite{pre93}) at $z=3$.  In the simulations,
changing the value of $h$, $\Omega_b$ or the ionizing flux results in
a rescaling of the optical depth.  To a good approximation
(particularly for the lower-column density lines) this simply has the
effect of multiplying the column densities by a constant factor and
shifting the distributions shown in Figure~\ref{fig:NH_dist}.  This
changes the number of lines observed in a given interval in $N_{HI}$,
however, it has relatively little effect on the shape of the
$b$-distribution or on its median value, which changes by only 1.5
km/s 
if we renormalize to a mean
flux decrement of 0.36 (as opposed to values of around 0.22 without
renormalization).

\subsection{What causes $b$ to decline with resolution?}
\label{sec:b_decline}

The width of a particular line can be affected by the gas temperature,
the peculiar velocity structure, and the density profile of the line.  An
increase in the temperature would increase the Doppler broadening; an
expanding structure (or even one collapsing less quickly) can broaden
the profile; finally the density distribution itself can give rise
to such an effect since a line that is physically wider will cover
more of the Hubble flow.

\begin{figure}
\epsfxsize=3.5in
\centerline{\epsfbox{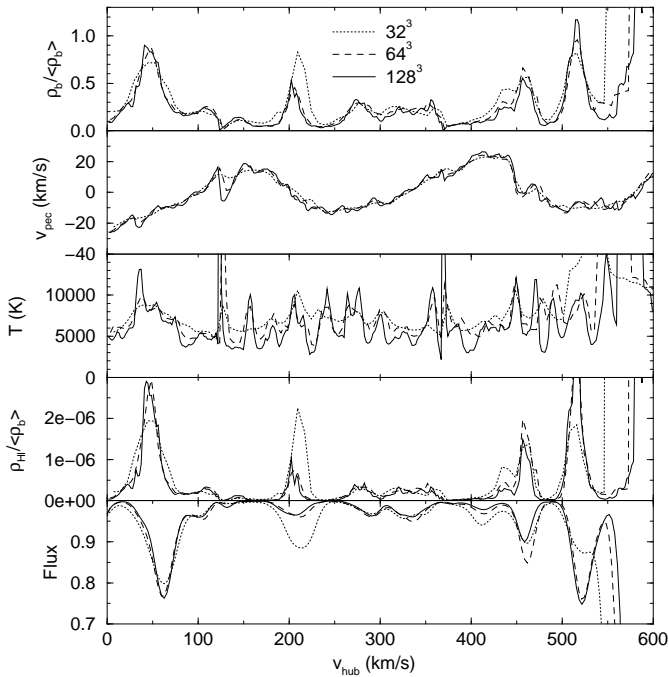}}
\caption{
The profiles of various quantities resulting from the same
line-of-sight taken through the three simulations at $z=3$.
The ordinate is the Hubble velocity corresponding to that cell (for
baryonic density, temperature, velocity, and HI density) or the
wavelength (converted to km/s) for the flux.
}
\label{fig:flos_5panel_z3}
\end{figure}

In order to determine the cause of the decline seen in the previous
section, we must look at the underlying gas distribution that gives
rise to each line.  In Figure~\ref{fig:flos_5panel_z3} we plot the
same line-of-sight for the three different resolutions at $z=3$.
There is rough correspondence between large features, however the
details vary strongly, particularly for the $32^3$ run (compared to
the others).  Unfortunately, a comparison of this type is not
particularly useful due to the chaotic nature of the non-linear
evolution; slight differences in, say, the position of a clump can
give rise to large discrepancies along a particular line of sight.
Still, a number of broad conclusions can be drawn.  The first is that
the large-scale velocity and density structures are well followed,
although in at least one case (around 200 km/s), the differences are
surprisingly large.  The two better-resolved simulations tend to track
each other with satisfying fidelity.  The second is that a
qualitatively new feature appears to have arisen in the
high-resolution temperature profiles --- high amplitude fluctuations
on the 20 km/s (100 $h^{-1}$ kpc) scale.  We will return to this
observation below.

While such anecdotal evidence is interesting, it is more useful to
look at the mean behaviour of the lines.  The difficulty in doing this
is to associate a given flux decrement (line) with the range
of cells which create it.  Because of Doppler shifts due to peculiar
velocities in the gas, there does not have to be a one-to-one
correspondence.  However, for the vast majority of lines, the
correspondence is clear.
Still, we need to remove the peculiar velocity contribution in order
to do this association; fortunately, the bulk velocity component
changes slowly, so this task is relatively straightforward.


Another way to state the problem is that given the center of a line,
$v_{line}$, in the optical depth distribution $\tau(v_{line})$, how do
we find the corresponding point $x$ in $\rho(x)$ or $T(x)$, where $x$
is the distance along the line of sight?  The two are related through
$v_{line} = H(z) x + v_{pec}(x)$, or more precisely, $\tau(v_{line})$
is given by an integral over $x$, which we write schematically as
$\tau(v_{line}) = \int (d\tau/dx) dx$ (for more details see
\cite{zha97b}).  Since the $v_{line}$-$x$ relation cannot be directly
inverted, the obvious approach is to solve it iteratively; however,
this runs into convergence problems.  Instead, we replace the term
$v_{pec}(x)$ with:
\begin{equation}
	\bar{v}_{pec}(v_{line}) = \tau(v_{line})^{-1}
\int{\frac{d\tau}{dx} v_{pec}(x) dx},
\end{equation}
which is the (optical-depth weighted) mean peculiar velocity of the
absorbing gas.  This value is then used to correct the observed
line-center to our estimate of the line-center as measured by the
Hubble velocity along the line-of-sight: $v_h = H(z) x = v_{line} -
\bar{v}_{pec}$.

\begin{figure*}
\epsfxsize=7.0in
\centerline{\epsfbox{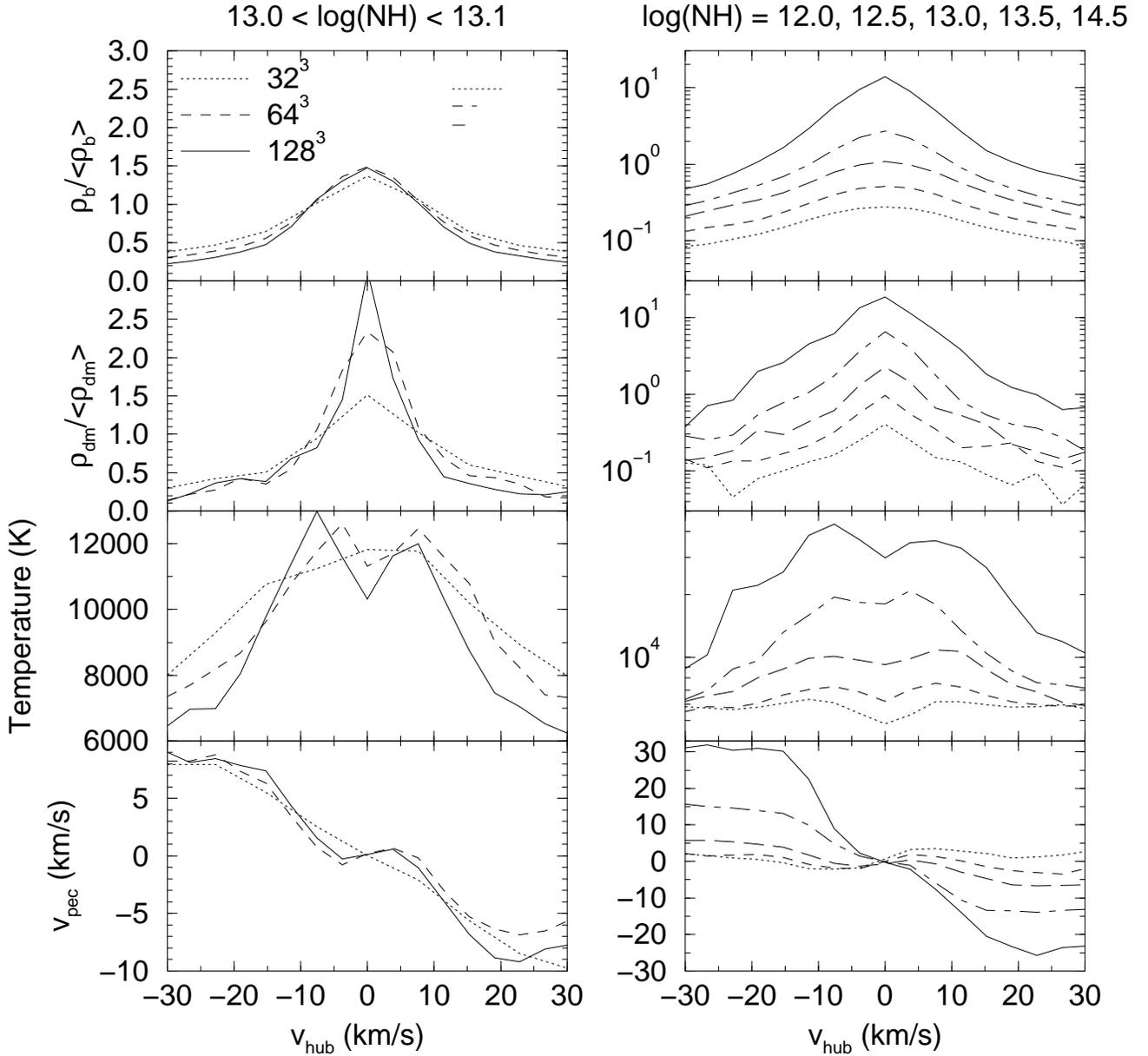}}
\caption{ Mean profiles, at $z=3$ of (from top to bottom) baryon
overdensity, dark matter overdensity, temperature, and peculiar
velocity projected along the line of sight.  The left column shows
results from three simulations with varying resolution for lines with
neutral hydrogen column densities in the ranges $N_{HI} =
10^{13.0}-10^{13.1}$ cm$^{-2}$.  In the upper left panel, three lines
show the cell size for each simulation.  The right column shows the
same quantities from the highest resolution simulation for a range of
column densities.  From bottom (dotted) to top (solid), the lines
correspond to column densities in the ranges: $N_{HI} =
10^{12.0}-10^{12.1}, 10^{12.5}-10^{12.6}, 10^{13.0}-10^{13.1},
10^{13.5}-10^{13.6}$ and $10^{14.5}-10^{14.7}$ cm$^{-2}$.  }
\label{fig:mean_line1}
\end{figure*}

We then follow the LOS in both directions (positive and negative
velocity) until a density minimum is encountered in the baryon
distribution along each path.  The cells between these two density
minima are extracted and used to compute the mean profiles shown in
Figure~\ref{fig:mean_line1} (in order to prevent small fluctuations
from introducing small minima which would cause us to extract too
small a section, we smooth the density slightly, but only while
checking for minima).  We discard lines that do not have a density
maximum within 20 km/s of the estimated line center (about 5\% of the
total), since visual inspection of a sample of these showed them to
fall on the wings of much larger lines.  This entire procedure is
robust to small changes in the details.


Since the column density is a robust prediction, we extract those
lines which lie within a small range of $N_{HI}$ and plot the results
in Figure~\ref{fig:mean_line1}.  We first discuss the left hand
column, which shows a resolution study for a large number of lines
(drawn from 300 samples each of length $\Delta z = 0.1$) from the same
three simulations discussed previously.  Only lines in the range
$N_{HI} = 10^{13.0}$--$10^{13.1}$ are used.  Although individual lines
can differ in detail, these mean profiles are characteristic; to give
an idea of the variation, the {\it rms} scatter in
$\rho_b/\left<\rho_b\right>$ is about 0.3.  The dark matter, which is
represented as particles rather than cells in the simulation, is
interpolated to a mesh with the tri-linear cloud-in-cell method.
We first address the effect of resolution, and then, in
the next section, turn to what the figure can tell us more generally
about the lines.

Systematic resolution effects are visible in each quantity.  The
density profiles, both baryonic and dark, are narrower with higher
resolution, although the largest changes are in the dark matter
profile, while the baryons show a flatter central profile.  The
temperature profile is relatively unchanged, although it becomes
somewhat narrower and features a central depression at high
resolution.  The peculiar velocity profile develops a kink in the
central 20 km/s with cell sizes less than 37.5 $h^{-1}$ kpc.

\begin{figure}
\epsfxsize=3.5in
\centerline{\epsfbox{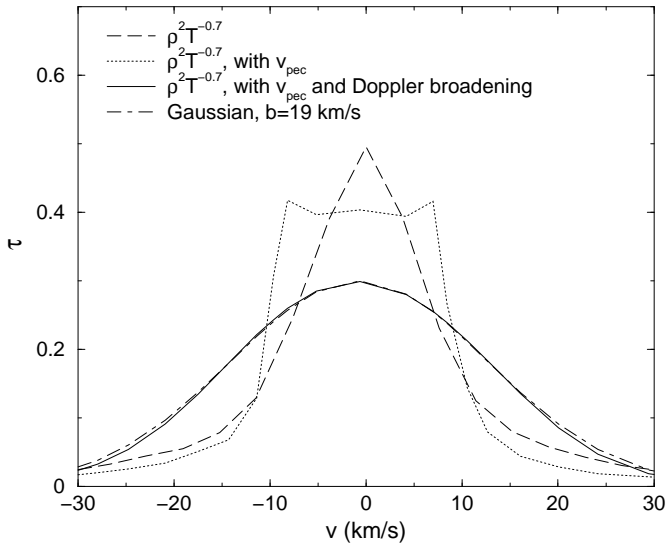}}
\caption{ This figure shows how Hubble broadening, peculiar velocities
and thermal broadening affect the mean line profile for lines with
$N_{HI}$ in the range $10^{13.0}-10^{13.1}$ cm$^{-2}$.  The dashed
line shows an estimate of the optical depth ($\rho^2 T^{-0.7}$, from
the $128^3$ simulation shown in Figure~\protect\ref{fig:mean_line1}),
the dotted line adds the distortion due to the mean peculiar velocity
profile, and the solid line also includes thermal broadening.  The
dot-dash lines demonstrates that a Gaussian
is a good fit to the resulting profile.  
}
\label{fig:log13_tau}
\end{figure}

This figure confirms (\cite{wei97}) that a significant part of the
line-broadening --- at least for low-column density lines --- is due
to the Hubble flow across the line width, although thermal Doppler
broadening also makes an important contribution.  These statements are
made a little more precise in Figure~\ref{fig:log13_tau}, which
converts the mean density and temperature profiles from
Figure~\ref{fig:mean_line1} into an optical depth, assuming ionization
equilibrium.  In this case the HI opacity is proportional to $\rho^2
\alpha(T) \sim \rho^2 T^{-0.7}$, where $\alpha(T)$ is the
recombination coefficient.

Without any peculiar velocities or thermal broadening, the resulting
profile would be the dashed line.  For such low optical depths, a
Voigt profile is indistinguishable from a Gaussian.  Although
approximately Gaussian in shape, the profile shown here is not exactly
so, and fits produce $b$ parameters ranging from 10 to 15 km/s,
depending on which part of the curve is fit.  The dotted line includes
the mean peculiar velocity, which shifts the observed velocities of
the profile ($v = v_H + v_{pec}$), and multiplies the optical depth by
$(1 + dv_{pec}/H(z)dx)^{-1}$. The boxy shape is a result of the kink
in the velocity profile.  Adding thermal broadening is enough to make
the spectrum quite Gaussian, as the fit in the figure demonstrates.
This is important, since it is this shape that is used to fit lines in
the observed spectrum and suggests that Voigt-profile fitting,
although often criticized, may be more robust than commonly assumed.
Still, it must be kept in mind that in this exercise we have averaged
the density, temperature and peculiar velocity separately and then
formed a mean profile, which is quite different from averaging the
optical depth distribution directly.  Moreover, even if the average
shape is Gaussian, individual lines may vary substantially.

Returning to Figure~\ref{fig:mean_line1}, we can now try to answer the
question posed earlier.  Of the three possible causes for the decrease
in the line width, the most significant is the density, which shows a
clear decrease in width as the resolution improves.  This is
particularly true in the outskirts of the line, which contributes
preferentially to the line thickness.  This can be roughly quantified
by the width of a Gaussian fit to the density profile.  Although it is
approximately Gaussian in shape, it is impossible to fit both the
inner and outer parts with the same parameters; however, using only
the outer points (beyond $\pm$ 10 km/s), the fit width decreases by
25\% in going from the $32^3$ to the $128^3$ simulation.  The inner
part shows a less dramatic 10\% decline.  These figures are comparable
to the drop in the median $b$ with resolution.

This plot also shows that the change in the peculiar velocity is not
primarily responsible for increasing the line width, since the
magnitude of the change is only around 1 km/s.
The temperature also adds to the width somewhat, as the
high resolution profile is narrower, so the gas temperature in the
line edges is reduced.  This reduces the Doppler broadening (which
contributes a width of approximately 13 $(T/10^4 $K$)^{1/2}$ km/s),
but again only by about 1 km/s or so.

From this discussion, we can conclude that the primary reason for the
shift in the $b$ distribution is numerical thickening of the lines.
This shows up most clearly in the density profile, but it also affects
the temperature distribution.




\subsection{The physical nature of the absorbers}
\label{sec:nature}

These profiles also tell us a great deal about the physical nature of
the absorbers.  The velocity plot (still focusing on the left hand
side of Figure~\ref{fig:mean_line1}) indicates that these lines are
infalling in comoving coordinates, but the amplitude, which is less
than the Hubble expansion, reveals that the objects are actually
expanding in absolute coordinates.  Thus, while the overdensity
($\rho_b/\left<\rho_b\right>$) may be increasing slowly, the absolute
density is dropping (at least for absorbers with this column density).

The temperature follows largely from the density profile: the
recombination rate is larger at higher densities and so
photoionization heating is more effective.  Lower densities have also
undergone more expansion cooling.  This dependence on density results
in a polytropic equation of state (\cite{hui97b}; see also
\cite{gir96}): $T \sim \rho^{0.4}$, by $z = 3$.  The temperature
contributes a width to the lines via Doppler broadening which is
approximately 13 $(T/10^4$K$)^{1/2}$ km/s, and so dominates the width
only for lines which are intrinsically thin.

Finally, we turn to the details of the line center.  The baryon
density profile shows a rounded top as opposed to the cuspy center of
the dark matter.  This is due to the effect of thermal pressure.
Prior to reionization, the gas and dark matter profiles follow each
other; however, once reionization occurs the gas temperature is raised
everywhere to a nearly constant 15,000 K (the exact temperature
depends somewhat on the assumed ionizing spectrum).  The gas in the
center of these objects, which are most likely to be sheets or
filaments (e.g. \cite{zha97b}), finds itself over-pressured and starts
to expand.  This (relative) expansion can be seen in the peculiar
velocity profile as the kink mentioned previously, and also cools the
gas, causing the central temperature depression.

To see that the pressure gradient is sufficient to accomplish this, we
note that the acceleration from the pressure $p$ is $\nabla p / \rho
\approx c_s^2\nabla \rho / \gamma \rho$, where $c_s$ is the sound
speed and the second step uses the fact that the temperature is
nearly constant at recombination.  Approximating the gradient
$\nabla \rho$ as $\Delta \rho / \Delta r = \Delta \rho H(z) / \Delta
v$, we can write the velocity change after a Hubble time as $\Delta v
= c_s (\Delta \rho / \gamma \rho)^{1/2}$.  Since the sound speed is
about 10 km/s at $10^4$ K, and the gradient term is of order unity or
higher, it's clear that a velocity change of the required magnitude
can be generated
as long as the initial peculiar velocities are not too large.

On the right-hand side of Figure~\ref{fig:mean_line1} are plotted mean
profiles for a range of column densities from $N_{HI} = 10^{12}$ to
$10^{14.5}$ cm$^{-2}$ (now using only the highest resolution
simulation).  Above this value there are too few lines, with too much
diversity for a reasonable mean profile to be constructed.  Other
studies (\cite{zha97b}; \cite{cen97}) have shown that the highest
density absorbers (
$N_{HI} \ga 10^{14}$ at $z=3$ in this model) tend to be
quasi-spherical, lying at the intersection of filaments, while
progressively lower density absorbers occur in filaments and then
sheets.  We focus mostly on low column-density lines and so on these
latter two kinds of absorbers.  We remind the reader that this
correspondence between overdensity and column density depends not only
on redshift, but also on a number of model-dependent factors,
including $\Omega_b$, $h$ and the photo-ionization rate.  For example,
lines with column densities of $10^{14}$ cm$^{-2}$ are associated with
mean overdensities of 3-4 in these simulations, but 1-2 in those of
Dav\'e \etal (1998b)\markcite{dav98b}, primarily due to their
renormalization to a higher flux decrement.

The density profiles, particularly for the dark matter, are remarkably
similar in shape, showing little variation in width and are simply
scaled in amplitude.  The gas distribution becomes more and more
peaked for lines with larger $N_{HI}$, as the thermal pressure that
gives rise to the rounded distribution for low-column density lines
becomes less important (see also \cite{mei94}).  This can be
quantified by comparing the FWHM of the dark matter and gas profiles,
as is done in Table~\ref{table:widths}.  The ratio varies from 3:1 at
low column densities, to nearly equal by $N_{HI} = 10^{14.5}$
cm$^{-2}$.


{
\begin{deluxetable}{cccccc}
\footnotesize
\tablecaption{Relative contributions from Hubble and thermal
broadening to lines of various column densities}
\tablewidth{0pt}
\tablehead{
\colhead{} & 
\colhead{$N_{HI}=10^{12.0}$} & \colhead{$N_{HI}=10^{12.5}$} &
\colhead{$N_{HI}=10^{13.0}$} & \colhead{$N_{HI}=10^{13.5}$} & 
\colhead{$N_{HI}=10^{14.5}$}
}
\startdata
peak overdensity & 0.28 & 0.51 & 1.1 & 2.7 & 13 \nl
0.46FWHM$_{dm}$   &  4.9 &  4.2 &  4.2 &  4.1 &  4.8 \nl
0.46FWHM$_{gas}$  & 14.8 & 12.4 & 10.6 &  7.5 &  5.2 \nl
$b_T$ (thermal) &  9.7 & 10.5 & 12.4 & 16.8 & 23.3 \nl
$\sqrt{(\hbox{0.46FHWM$_{gas}$)}^2 + b_T^2}$
                & 17.7 & 16.2 & 16.3 & 18.4 & 23.9 \nl
median $b$ from fits & 14.9 & 15.6 & 18.0 & 20.1 & 22.6 \nl
\enddata
\label{table:widths}
\end{deluxetable}
}


We can also use this information to obtain a rough estimate of the
relative contributions of Hubble and thermal broadening.  Since the
optical depth is given by $\tau \sim \rho^2 T^{-0.7}$ and $T \sim
\rho^{0.4}$ (see Figure~\ref{fig:hist2d} and \cite{hui97b}), we can
write $\tau \sim \rho^{1.7}$.  If the density distribution were a
Gaussian, then the optical depth profile would also be a Gaussian,
with $b = $FWHM$/(2\sqrt{1.7\ln2}) \approx 0.46$ FHWM.  In
Table~\ref{table:widths} we give this value, which is roughly the
contribution from Hubble broadening to the total width of the line.
We also include the $b$ parameter that would result from thermal
broadening alone, using the mean temperature of the profile (weighted
by $\rho^2 T^{-0.7}$).  In the second last row, these two values
are added in quadrature.  The result is 
broadly consistent with the median $b$ derived from fitting Voigt
profiles to the spectra, which is shown in the final row.  The
difference is most likely because 0.46FWHM is an underestimate of the
effect of Hubble broadening, and it should be kept in mind that we are
ignoring the effect of peculiar velocities.  Note in particular that
as the Hubble width of the lines decrease with increasing column
density, the temperature grows, leaving the resulting width
more-or-less constant.  This is consistent with the lack of a strong
correlation between $N_{HI}$ and $b$ in the simulations, and helps to
explain why the median of the $b$ distribution is insensitive to the
normalization of the mean decrement.

The central temperature dip and the velocity kink become less
pronounced with increasing column density up until $N_{HI} \sim
10^{14}$.  At this point, the peculiar infall velocity actually
becomes larger than the Hubble flow and the object has detached from
the Hubble expansion and is collapsing.  Once this happens, two shocks
form on either side of the mid-plane and propagate outwards (visible
in the velocity distribution at $\pm 10$ km/s).  If this occurs in a
sheet, then the object forms a Zel'dovich pancake (otherwise the shock
is cylindrical in the case of a filament or quasi-spherical for a
knot).  These lines are now dominated by thermal broadening, since
$v_{pec} \sim -v_{hub}$.  It is interesting to note that this
qualitative change in the physical nature of the line occurs at
$N_{HI} \sim 10^{14.5}$ cm$^{-2}$, which is also where it appears that
the carbon abundance (and by plausible extension, all metals)
undergoes a drastic change: from [C/H] $\sim -2.0$ above this cutoff,
to less than -3.5 below (\cite{lu98}).  Similar results have been
found using OVI as a probe (\cite{dav98}).  This behaviour is in broad
agreement with that predicted in the simulations of Gnedin \& Ostriker
(1997)\markcite{gne_ost97}, which explicitly include star formation
and the production of metals.

Finally, we comment on the fact that the density away from the line
drops below the cosmic mean, and in fact, the central values of the
smallest lines are consistently below this value.  This means that
most low column density lines are in voids.  The size of these regions
are smaller than familiar galactic voids and so the name mini-void has
been proposed (\cite{zha97b}).

\subsection{Resolving the Jeans length: reionization-driven winds}
\label{sec:jeans}

\begin{figure}
\epsfxsize=3.4in
\centerline{\hspace{-0.3cm}\epsfbox{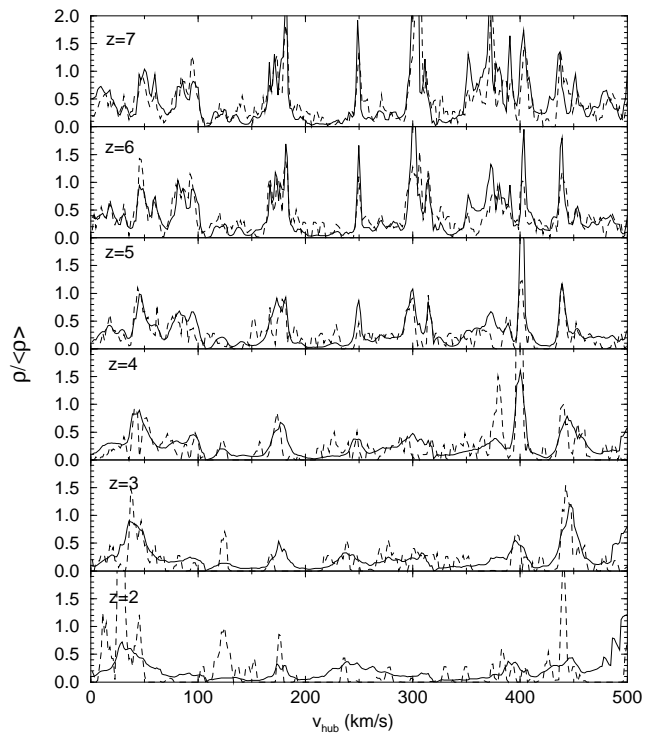}}
\caption{ The evolution of the baryon (solid) and dark matter (dashed)
overdensities 
($\rho/\left<\rho\right>$) along a line of sight for the highest
resolution simulation ($128^3$ cells).  The ordinate is the Hubble
velocity at $z=2$, with the other redshift ranges adjusted to cover
the same comoving length.  }
\label{fig:flos_dens128}
\end{figure}

\begin{figure}
\epsfxsize=3.6in
\centerline{\hspace{-0.3cm}\epsfbox{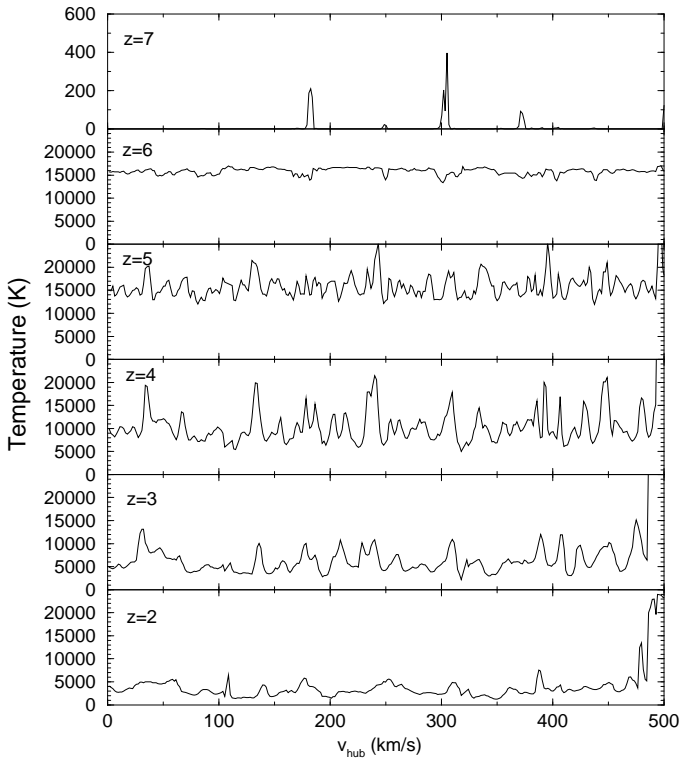}}
\caption{
The evolution of the baryon temperature along the same line of
sight shown in Figure~\protect\ref{fig:flos_dens128}.
}
\label{fig:flos_temp128}
\end{figure}

\begin{figure}
\epsfxsize=3.5in
\centerline{\hspace{-0.3cm}\epsfbox{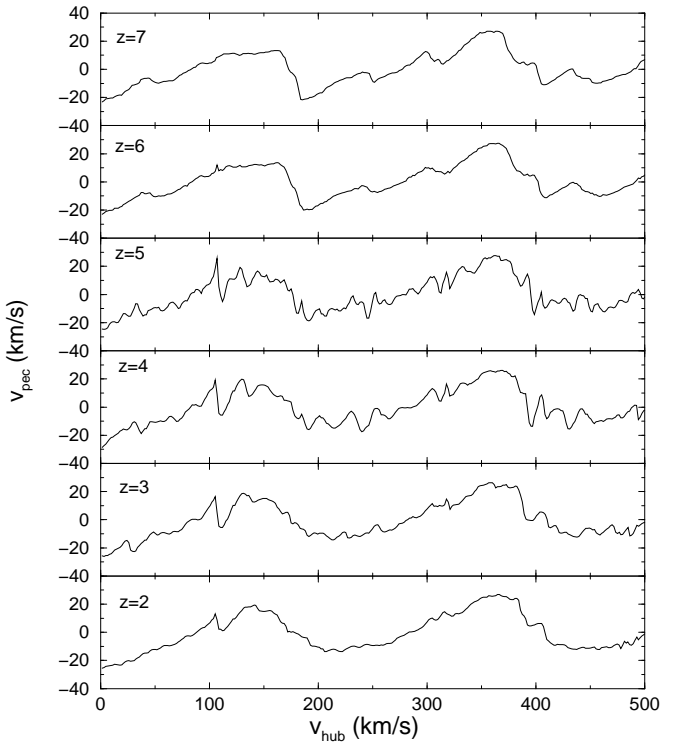}}
\caption{
The evolution of the projected baryon peculiar velocity along the same
line of sight shown in Figure~\protect\ref{fig:flos_dens128}.
}
\label{fig:flos_vproj128}
\end{figure}

The thermal pressure becomes more important for smaller lines until it
is sufficient to disperse the feature entirely (\cite{bon88}).  As we
pointed out earlier, the highest resolution simulations show small
scale fluctuations in the velocity and temperature distributions that
are mostly absent in the lower resolution runs.  Here, we offer an
explanation for that behaviour.  In Figures~\ref{fig:flos_dens128},
\ref{fig:flos_temp128}, and \ref{fig:flos_vproj128}, we plot the
baryon and dark matter density, the temperature and the projected
peculiar velocity along a line-of-sight of constant comoving
length at redshifts from $z=7$ to $z=2$.

At the earliest time, many of the smallest clumps (which are not fully
resolved even at this resolution) have collapsed into sharp density
spikes, and the baryons follow the dark matter distribution quite
well.  In this model, re-ionization occurs shortly after $z=7$ and by
$z=6$, the gas is almost all uniformly at $T=15,000$ K, regardless of
density.  The thermal pressure is then proportional to the density
(since $p \propto \rho T$) and the clumps, filaments and sheets which
are not sufficiently massive to generate a gravitational force larger
than this thermal one begin to expand.  Roughly, this constraint can
be expressed as the requirement that an object must be larger than the
Jeans length: $\Delta v_j = H(z) r_j = 26.3$ $(T/10^4 K)^{1/2}$ km/s.

After re-ionization, the smallest objects expand, driving out a wind.
The result can be seen in the density field, as a smoothing of the
smallest scales, but it is in the temperature and velocity fields
where the changes are the most startling.  By $z=5$, these winds
produce small-scale fluctuations in the velocity, and in the
temperature, as the resulting expansions and contractions cause
adiabatic cooling and heating.  The fluctuations decay at late times,
as the gas rearranges itself to new equilibrium configurations.  

We argued in section~\ref{sec:nature} that this process can accelerate
gas to velocities around the sound speed for the lowest column density
lines, so for these objects -- which are smaller -- the gas velocities
should be slightly lower.  Indeed, Figure~\ref{fig:flos_vproj128}
shows fluctuations of around 5 km/s.  For an adiabatic gas with a
ratio of specific heats $\gamma$, the temperature change will be
$\Delta T/T = (\gamma-1)\Delta \rho/\rho$.  From
Figure~\ref{fig:flos_dens128}, we see that $\Delta \rho/\rho$ can
approach unity, so large changes in the temperature are expected
provided that the timescale for radiative processes is longer than the
dynamical time (i.e. the gas is adiabatic).  The dominant heating
mechanism is photo-ionization heating, which is limited by the
recombination time-scale: $t_R = 2.3 \times 10^{17} (T/10^4 K)^{0.7}
((1+z)/4)^{-3} (\rho/\bar{\rho})^{-1}$ s, while a 5 km/s wind can cross
two cells (approximately our resolution limit) in $5.8 \times 10^{16}
((1+z)/4)^{-1}$ s.

It is these winds, driven by reionization, that produce the scatter
in Figure~\ref{fig:hist2d}, as the resolution improves.  In detail,
the process of dispersing sub-Jeans length clouds is not modeled
correctly, partially because we do not have the necessary resolution,
but also because we have approximated the radiation field as uniform
and isotropic.  In fact, discrete ionizing sources will produce
I-fronts, which will drive a rocket effect, accelerating the cloud in
the direction of the radiation, until it is completely ionized and
disperses (\cite{sha98}).  Still, the net result should be grossly
similar unless an object is sufficiently dense that it is optically
thick to the ionizing radiation.  Such clouds, in order to survive,
will have high column densities ($N_{HI} > 10^{18}$) and have been
proposed as the source of some Lyman-limit systems (\cite{abe98}).
However, unless a significant fraction of the mass is tied up in such
systems, they will not greatly affect our results.



\subsection{Other statistical measures}
\label{sec:non_param}

The objects that give rise to absorption features in the Ly$\alpha$
forest are generally not the thermally broadened, compact clouds that
the Voigt profile assumes (although it is interesting to note that
their mean shape is not so different from a Gaussian).  Since this
profile is not always a good fit to the observed (or simulated)
features, it makes sense to develop other, less parametric ways to
characterize the observations.  Perhaps the most straightforward is to
work directly with the N-point distribution functions of the
transmitted flux ($F = e^{-\tau_{HI}}$) itself (\cite{mir97};
\cite{rau97}; \cite{cen97b}). 

\begin{figure}
\epsfxsize=3.4in
\centerline{\epsfbox{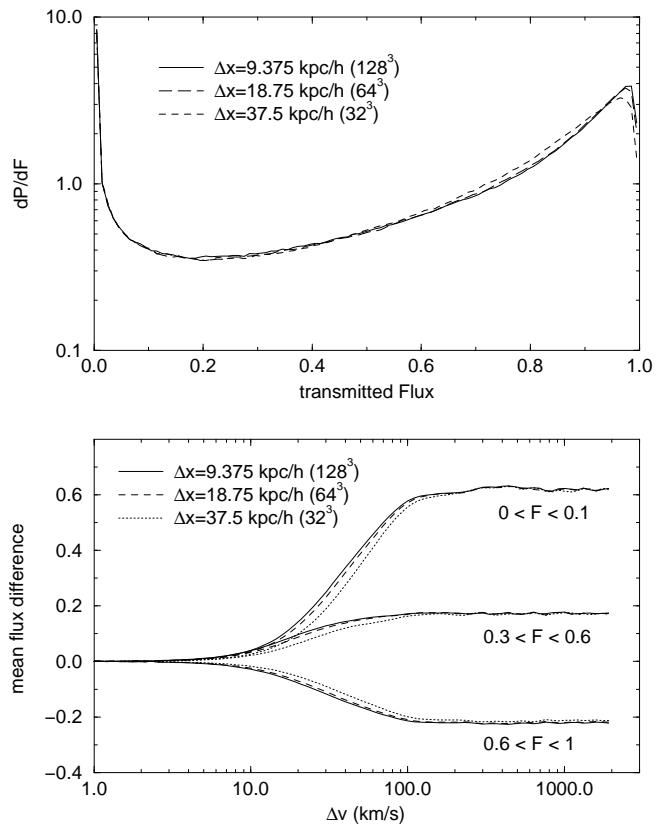}}
\caption{ The probability distribution (top) of the transmitted flux
at $z=3$ for the same three simulations with different resolutions
shown in Figure~\protect\ref{fig:NH_dist}.  The bottom panel plots the mean
flux difference as a function of velocity difference for pixels in the
ranges indicated (eq.~\protect\ref{eq:f2}).  }
\label{fig:moments}
\end{figure}

In Figure~\ref{fig:moments}, we show the probability distribution at
$z=3$ from 300 lines-of-sight through the same three simulations
previously discussed.  The photoionizing flux has been rescaled so
that the mean flux decrement ($D_A = 1 - F$) is 0.36 (\cite{pre93}).
The two higher-resolution results show
very similar distributions, while the lowest exhibits a deficit around
$F \sim 0.95$ and a corresponding surplus near $F \sim 0.8$.  This
follows simply from the profile broadening noticed earlier: narrow
lines dip down deeper and so produce lower values of the transmitted
flux (higher decrements), leaving more of the spectrum near its
continuum value.

This redistribution of the flux will obviously affect zero-point
statistics such as the mean decrement.  For example, without
renormalizing, the average decrement decreases with resolution: $D_A
= 0.229, 0.217, 0.214$ for the $32^3$, $64^3$ and $128^3$
simulations, respectively.  While this is relatively slight, the
effect on the He will be greater (see also \cite{zha97b}).

A higher order statistic than the flux distribution is the two-point
function $P_2(F_1,F_2,\Delta v)$, which gives the probability that two
pixels with separation $\Delta v$ will have flux $F_1$ and $F_2$.  In
order to present this function graphically, we plot normalized moments
of it, averaged over a range of flux ($F_a$ to $F_b$):
\begin{equation}
\label{eq:f2}
{
{\int_{F_a}^{F_b} dF_1 \int_0^1 dF_2 P_2(F_1,F_2,\Delta v)(F_1-F_2)}
\over
{\int_{F_a}^{F_b} dF_1 \int_0^1 dF_2 P_2(F_1,F_2,\Delta v)}
}
\end{equation}

This is the average flux difference as a function of velocity for
pixels in the range $F_a$ to $F_b$, and is plotted in the bottom panel
of Figure~\ref{fig:moments} for three ranges.  Although somewhat
complicated, it can be interpreted in the following way.  For small
separations, there is little difference regardless of the flux value,
or in other words there is high coherence for small $\Delta v$, a fact
which is clear from just looking at the spectra.  At large separations
($\Delta v > 200$ km/s), there is no coherence, and the value is just
the difference between the mean value of the transmitted flux (0.65)
and the mean flux in the interval $F_a$ to $F_b$.  For intermediate
separations, in the range $\Delta v = 10$ to 100 km/s, the plot shows
a measure of the mean profile around pixels with fluxes in that
interval (this is heavily influenced by the structure of lines, but
around 100 km/s line-line correlations also play a role).

The effect of resolution is again clear, narrowing the line profile as
the cell size decreases.  Therefore we see that this effect shows up
not just in the parametric analysis (Voigt-fitting), but in the
distribution of fluxes as well.  Although the difference looks smaller
in this plot, that is largely a function of the compressed scale.  For
example, if we take the $0 < F < 0.1$ profiles and ask at what $\Delta
v$ do the lines go through a mean flux difference of 0.3 (which is an
approximate measure of the line width), the answer is 43 km/s for the
$32^3$ simulation, but only 35 km/s for the $128^3$ run.  While the
absolute numbers differ, this records a drop in width of nearly 19\%,
comparable to the decline in the median $b$ parameter.


\subsection{The effect of large-scale power}
\label{sec:power}

In this section, we study the effect of additional large-scale power
(see also \cite{wad97}), to address the question: are the results of
the previous discussion biased by the small box size?

\begin{figure}
\epsfxsize=3.5in
\centerline{\epsfbox{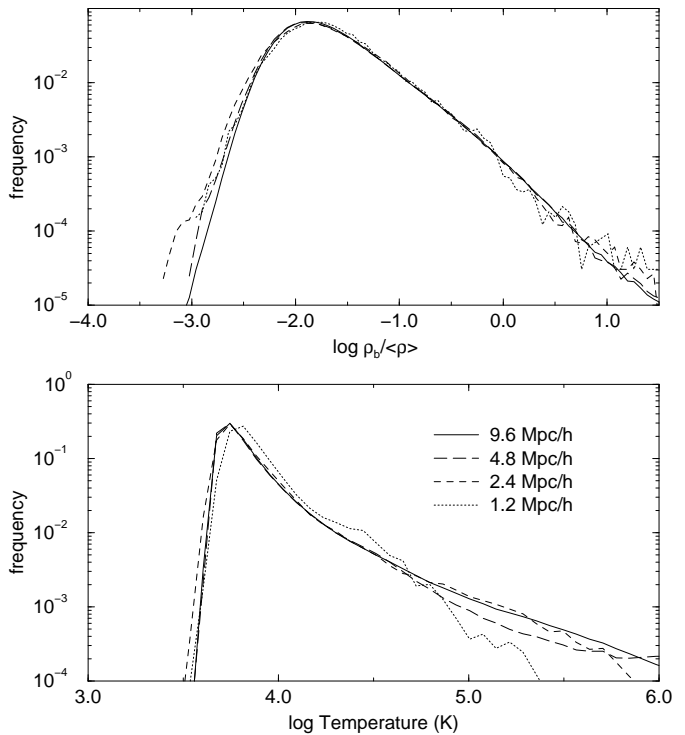}}
\caption{
The distribution of baryon overdensity (top) and volume-weighted
temperature (bottom) 
for four simulations with increasing box size --- but the same
resolution --- at $z=3$.
}
\label{fig:lss_hist}
\end{figure}

In Figure~\ref{fig:lss_hist}, we show density and (volume-weighted)
temperature distributions for four simulations with increasing amounts
of large-scale power but the same resolution.  The smallest box is the
same $32^3$ simulation analyzed above, while the largest is a $256^3$
simulation with a 9.6$h^{-1}$ Mpc box size.  The distributions are
remarkably similar, although there are some differences, particularly
for the two smallest boxes.

The biggest difference due solely to the presence of long-wavelength
modes is the increased amount of high-temperature gas.  As larger
modes are included, the bulk velocities increase and the gas that is
shock-heated around the edge of the biggest sheets, filaments and
knots is correspondingly hotter.  If we were looking at the formation
of very dense, and hence very rare, clumps (galaxies and groups), we
would also see an increase in the number of these objects; however,
their absence has little effect on the low-density regions that occupy
most of the volume and hence on the Ly$\alpha$ forest itself.  The
scatter at high density for the small boxes is simply due to Poisson
fluctuations since they have much smaller samples.

The other difference that is apparent is a small shift in the peak
temperature, and a slight rearrangement in the low-density end of the
baryon distribution (particularly evident in the low $\rho_b$ tail).
The evolution of the small boxes is mostly guided by the largest
modes in the box, which are not well sampled (i.e. are determined by a
small number of random values).  They become quasi-linear by $z \sim
3$ and mode-mode coupling starts to play a role, spreading their
influence over a wide range of scales.  Also, there
are no fluctuations above the box size so mode-mode coupling cannot
transfer that power down to smaller scales.  This means that, on average,
there tends to be a deficit of power in small boxes.

Put more simply, these boxes are not fair samples, In, fact, if the
smaller box sizes (1.2$h^{-1}$ and 2.4$h^{-1}$ Mpc) are repeated with
a different initial random seed, they show variations of the same
magnitude.  However, the two larger volumes show nearly identical
distributions.

\begin{figure}
\epsfxsize=3.5in
\centerline{\epsfbox{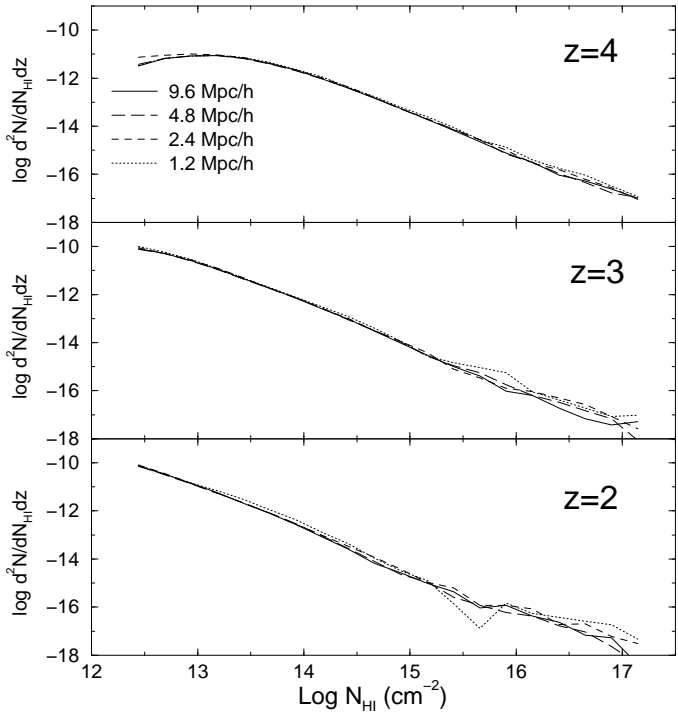}}
\caption{
The distribution of neutral hydrogen column densities for the same
simulations shown in Figure~\protect\ref{fig:lss_hist}, at three different
redshifts.
}
\label{fig:lss_nh_dist}
\end{figure}

Turning to the observable quantities, we show in
Figure~\ref{fig:lss_nh_dist} the distribution of column densities for
these same four simulations, demonstrating again the robustness of
this measure.  There are, however, some systematic differences,
notably for the smallest box which shows a distinct overproduction of
moderate to high column density lines.  This is again due to the fact
that we are not simulating an adequately large volume to contain a
reasonable sample, a situation which gets worse at lower redshift as
larger regions go non-linear.  Although in this case, the result is a
flattening of the power-law distribution, we argue that in general
either case could occur depending on whether the fluctuations are
systematically larger or smaller than average.  If we fit power laws
to these distributions, the slopes differ by less than 2\% for the two
largest boxes, but show variations of twice that value for the
smaller volumes.


\begin{figure}
\epsfxsize=3.5in
\centerline{\epsfbox{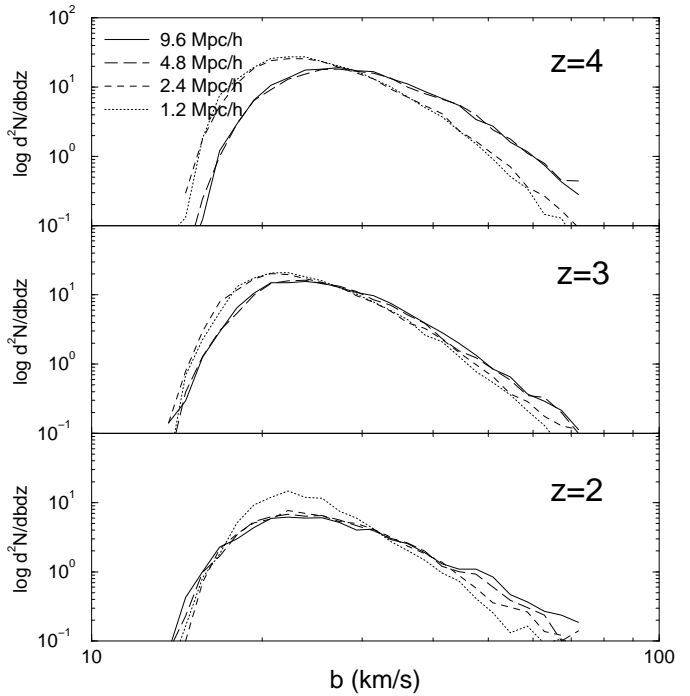}}
\caption{
The distribution of Doppler $b$-parameters for the same
simulations shown in Figure~\protect\ref{fig:lss_hist}, at three different
redshifts.
}
\label{fig:lss_b_dist}
\end{figure}

The same effect is operating in the distribution of Doppler $b$ parameters,
plotted in Figure~\ref{fig:lss_b_dist}, causing shifts in the entire
distribution for the smaller boxes, while the larger boxes show nearly
identical distributions.  
The magnitude is about 2.5 km/s at $z = 3$.

This is consistent with the fact that scales which give rise to
fluctuations on the few hundred kpc scale require box sizes of around
5 $h^{-1}$ Mpc to resolve properly.  For example, the measure of small
scale fluctuations suggested by Gnedin (1998)\markcite{gne97}:
\begin{equation}
\sigma^2_{34} = \int_{0}^{\infty} P(k) e^{-2k^2/k^2_{34}} 
{{k^2 dk} \over {2 \pi^2}}
\end{equation}
(where $k=2\pi / L$, $P(k)$ is the power spectrum at $z=3$ and 
$k_{34} = 34 \Omega_0^{1/2} h$ Mpc$^{-1}$) can be used to make this
argument more concrete.  If, instead of integrating from 0, we use $k =
L/2\pi$ with $L$ the size of the box, and denote this quantity
$\sigma^2_{34} (L)$, then the ratio of the power included within the
box to the total power: $\sigma^2_{34} (L)/\sigma^2_{34} =
0.67/0.82/0.92/0.96$ for $L = 1.2/2.4/4.8/9.6$ $h^{-1}$ Mpc.


\section{Discussion}
\label{sec:conclusions}

In the previous section we have shown that while the column density
distribution is not a sensitive function of resolution, the Doppler
$b$ distribution is: systematically shifting towards lower $b$ with
increasing resolution.  This is important because the median of this
distribution was already marginally low for this model, compared to
observations.  We demonstrated that this decrease is due to intrinsic
changes in the density profile, rather than the temperature or the
velocity structure of the lines.  The small volumes used to obtain
high resolution do cause some change in the median $b$, but not enough
to explain the discrepancy with observations.  In this section, we
turn first to modeling the effect of resolution, shedding some light
on the shape of the $b$-distribution.  Then we briefly discuss ways to
reconcile our result with observations.

\subsection{Modeling the $b$-distribution}

\begin{figure}
\epsfxsize=3.0in
\centerline{\epsfbox{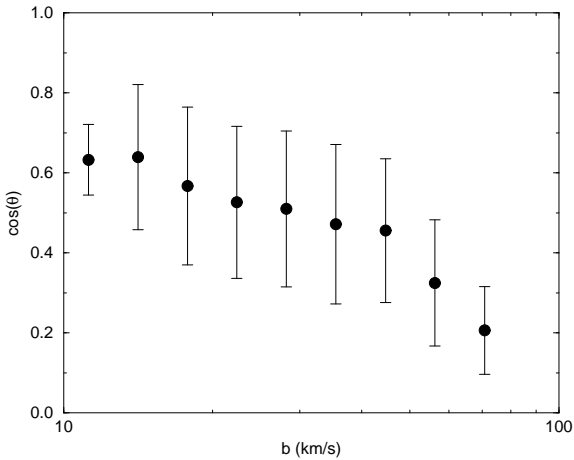}}
\caption{
The mean alignment of the density gradient and the line-of-sight
vector ($\cos(\theta)$) decreases with $b$.  In order to give an idea
of the scatter, the error bars show the standard deviation for each
bin.
}
\label{fig:b_costh}
\end{figure}

One of the traditional difficulties in understanding this distribution
is the presence of large $b$ values (e.g. \cite{pre93b}).  To be
explained by Doppler broadening, these systems require temperatures
too large to be explained by thermal equilibrium.  The simulations
also produce a large tail, and yet for large $b$ lines with a given
column density, we do not see a correlation between the width of a
line and the temperature of the underlying gas.  It has recently been
pointed out that such a tail is a natural consequence of the
filamentary and sheet-like nature of CDM-like structure formation
(\cite{rut98}), and the high-$b$ lines arise from oblique
lines-of-sight through these structures.  To test this idea, we
compute, for a large sample of lines, the alignment between the
density gradient and the line-of-sight vector:
\begin{equation}
\cos\left(\theta\right) = \sum{{{|\bf \nabla} \rho_b \cdot {\bf x}|} \over
{| {\bf \nabla} \rho_b |}}.
\end{equation}
The sum is over cells that give rise to the line and ${\bf x}$ is the
line-of-sight unit vector.  For sight lines which go perpendicularly
through a sheet or filament, the two vectors should line up perfectly
and $\cos(\theta) = 1$, while the measure goes to zero as ${\bf x}$
becomes parallel to the filament.  Figure~\ref{fig:b_costh} shows that
while there is a lot of noise in the relation, the correlation is
clear and goes in the expected direction.

This picture of the clouds also gives us insight into the effect of
resolution.  It has proved useful to model this effect in Eulerian
simulations as a Gaussian smoothing of the density field, with a
kernel of approximately two cells (\cite{bry98}).  Such a convolution
means that the width of a given structure --- whether sheet, filament
or sphere --- would be increased by adding the intrinsic width of the
structure and the size of the smoothing kernal in quadrature:
$l^\prime = \sqrt{l^2 + (2\Delta x/(1+z))^2}$, where $l$ is a measure
of the structure's intrinsic width in proper coordinates, $\Delta x$
is the comoving cell size and $l^\prime$ is the resolution-thickened
width.

If we now return to our model for the formation of the lines and make
two simplifying assumptions: (1) that all lines are due to the oblique
passage of a line-of-sight through filaments or sheets which all have
width $l$, and (2) that Hubble broadening dominates, we can construct
a method for correcting for this convolution.  Together these
assumptions imply that the width of a line is given by $b = H(z) l
\cos(\theta)$ and that the variation in $b$ is due to variation in
$\theta$, the angle between the line-of-sight and the normal of the
filament or sheet.  This means that the ratio of the numerically
thickened line width to its intrinsic width is given by:
\begin{equation}
f(\Delta x) = \frac{b^\prime}{b} = 
\left(1 + \left(\frac{2\Delta x}{l(1+z)}\right)^2\right)^{1/2}
\end{equation}
Notice that this expression does not depend on $b$, so it is a
constant for all lines (given our assumptions).  Unfortunately, we
don't know $l$; however, if we can associate $l$ with the width of the
median, $b_m$, then $l = b_m/H(z) = b^\prime_m f(\Delta x)/H(z)$ and
the correction factor $f(\Delta x)$ can be expressed directly in terms
of measurable quantities:
\begin{equation}
f(\Delta x) = \left(1 - \left(\frac{2\Delta x
H(z)}{b_m^\prime(1+z)}\right)^2\right)^{-1/2}.
\label{eq:b_correct}
\end{equation}




This provides a method to correct the measured $b$-parameter of a
given line which depends only on the measured median of the
(numerically thickened) distribution and $\Delta x$, the comoving cell
size (as well as some cosmological factors): we simply replace the
measured $b^\prime$ with $b = b^\prime /f(\Delta x)$.  Of course,
given our simplistic assumptions, it is unrealistic to expect this to
work on a line-by-line basis, but there is some hope it may provide a
good statistical correction.  We expect it to be least accurate for
lines arising from spherical absorbers, which tend to be higher column
density lines anyway, and for which our assumption of Hubble
broadening also probably fails.  This is one reason we restrict our
range of $b$ parameters to the lower column density end ($N_{HI} =
10^{13}-10^{14}$ cm$^{-2}$).  It will also be less accurate for those
lines which have small values of $b$ and are probably dominated by
variation in $l$ rather than $\theta$.

\begin{figure}
\epsfxsize=3.7in
\centerline{\epsfbox{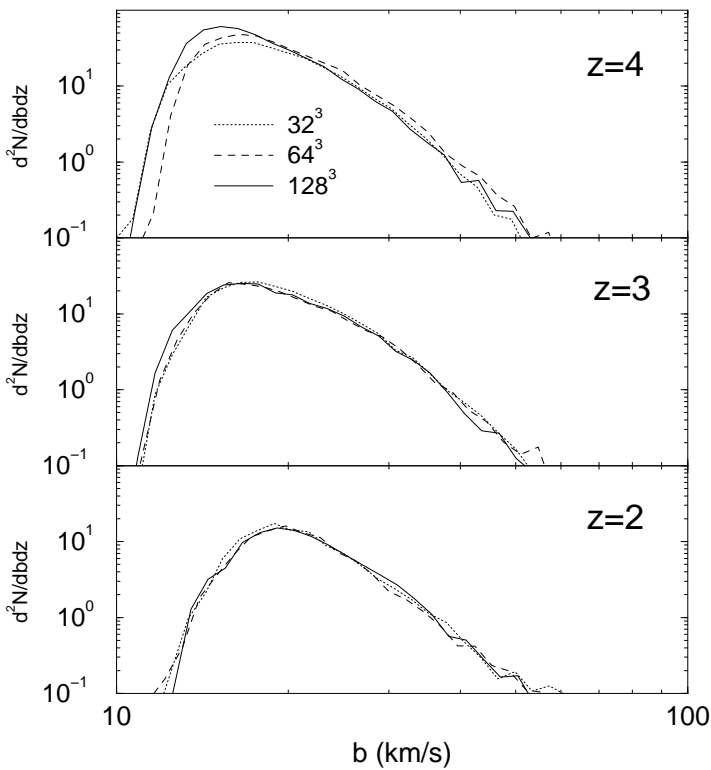}}
\caption{ The distribution of Doppler $b$ parameters (for lines with
column densities $10^{13} < N_{HI} < 10^{14}$ cm$^{-2}$) for the same
three redshifts and resolutions as in figure~\protect\ref{fig:NH_dist} (the
box size is 1.2$h^{-1}$ Mpc).  The distributions have been scaled to
account for resolution as described in the text.  }
\label{fig:b_log_13_14_scaled}
\end{figure}

We test the correction on the distributions plotted originally in
Figure~\ref{fig:b_dist}.  Since they are log-log plots, we can
simply shift the curves to the left (and up, since the distribution is
per unit $b$) by the constant factor $\log{f(\Delta x)}$.
The result is shown in Figure~\ref{fig:b_log_13_14_scaled}.

This scaling works remarkably well, except for the low $b$ end of the
distribution at $z=4$ (and to a lesser extent at $z=3$).  This is not
surprising, given that our approximation breaks down for low $b$
values.  Still, it provides a useful way to correct lower resolution
simulations: distributions can be scaled as described above and the
median (or mean) corrected by equation~\ref{eq:b_correct}.  For
example, the median values predicted at $z=2$ are 21.0, 21.3,
21.4 km/s for the the $32^3$, $64^3$ and $128^3$ simulations.  For
$z=3$ these numbers are 19.8, 19.7 and 19.5 km/s, while for the worst
case, $z=4$, they are 18.1, 18.9 and 17.5 km/s.

The ability to do this correction is important since it is difficult
to perform a simulation which is both large enough and of sufficient
resolution to reliably calculate the distribution.  We make a best
estimate for the median of the $b$ distribution in the range
$N_{HI}=10^{13}$-$10^{14}$ cm$^{-2}$ using two simulations: one (L9.6
from Table~\ref{table:simulations}), our largest simulation, but at a
relatively low resolution, and another (L4.8HR) with better
resolution, but a somewhat smaller box.  The results are shown in
Table~\ref{table:bmed}, along with the uncorrected values.


{
\begin{deluxetable}{cccc}
\footnotesize
\tablecaption{Best Estimate Median $b$-parameter}
\tablewidth{0pt}
\tablehead{
\colhead{simulation} & $z=2$ & $z=3$ & $z=4$
}
\startdata
L9.6 (corrected) & 24.8 & 22.3 & 25.8 \nl
L4.8HR (corrected) & 22.7 & 19.2 & 20.4 \nl
L9.6 (uncorrected) & 28.0 & 26.9 & 30.8 \nl
L4.8HR (uncorrected) & 23.6 & 20.7 & 22.1 \nl
\enddata
\label{table:bmed}
\end{deluxetable}
}


\subsection{The conflict with observations}

\begin{figure}
\epsfxsize=3.7in
\centerline{\epsfbox{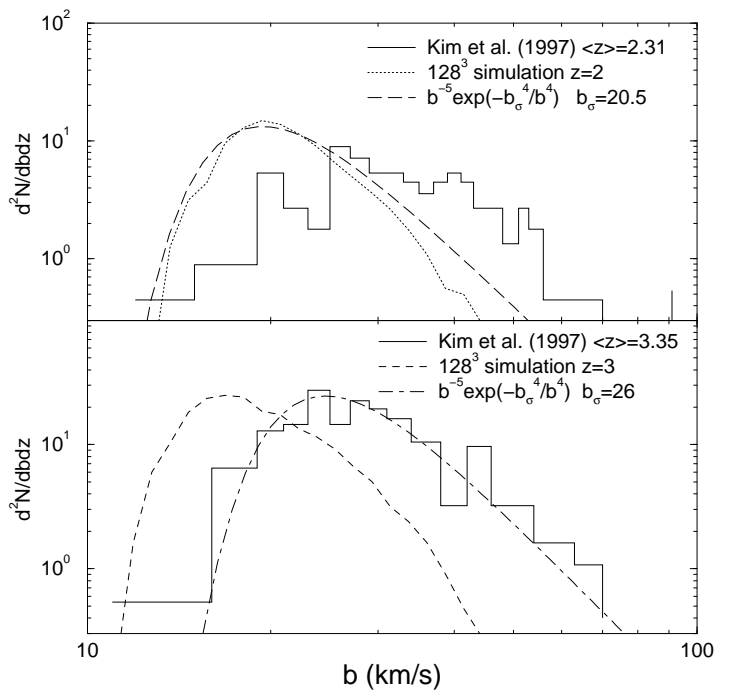}}
\caption{
The distribution of $b$ parameters for lines with column
densities $10^{13} < N_{HI} < 10^{14}$ cm$^{-2}$.  The top panel
shows observational data (\protect\cite{kim97}) at a mean redshift of 2.31 as
well as the $b$ distribution from our $128^3$ simulation at $z=2$;
The dashed line is an analytic profile fit to the simulated data.
The bottom panel shows the profile for observational data at $z=3.35$
and simulated data from $z=3$.  The same analytic profile is now fit
to the observations.
}
\label{fig:b_obs}
\end{figure}

To highlight the fact that the predictions of this model for the
$b$-parameter do disagree with observations, we plot both in
Figure~\ref{fig:b_obs}, for $z \sim 2$ and $z \sim 3$.  While the
shapes are roughly in agreement, the predicted medians are
substantially too low, in both cases (the observed median is about 30
km/s).  In this same plot, we also show the analytic profile suggested
by Hui \& Rutledge (1998)\markcite{hui98}, based on the Gaussian
nature of the underlying density and velocity fields:
\begin{equation}
\frac{dN}{db} \propto \frac{b_\sigma^4}{b^5} 
	\exp{\left[-\frac{b_\sigma^4}{b^4}\right]}.
\end{equation}
The normalization and the parameter $b_\sigma$ have been adjusted to
fit the simulations (observations) in the top (bottom) panel, in order
to demonstrate the relatively good agreement in shape.

We remind the reader that our analysis technique is not the same as
that used by observers, so we can not definitively rule out the
possibility that this discrepancy is due to our difference in fitting
Voigt-profiles to the spectrum (although we point out that Theuns
\etal (1998) find a similar discrepancy with yet another analysis
technique which is even closer to that used by the observers).  Still,
we have carried out a number of tests (see section~\ref{sec:lines})
which indicate that it is relatively easy to measure the line
width in our chosen column density range and that a bias of the
required magnitude would be difficult to arrange.  Probably the best
way to confirm this result would be a non-parametric test dealing
directly with the measured flux (as in section~\ref{sec:non_param}).
An early indication of which way this will go can be seen by comparing
our Figure~\ref{fig:moments} with its observationally determined
counterpart in Figure~3 of Miralda-Escud\'e \etal (1997).  By
determining where the $0 < F < 0.1$ curves go through a mean flux
difference of 0.3 in the respective plots, we can obtain a rough idea
of the difference in the widths: for the observations, this is $\Delta
v \approx 55$ km/s as compared to 35 km/s for our highest resolution
simulation.

Previous work on the Ly$\alpha$ forest has, more often than not,
stressed that the models were consistent with observations.  In fact,
this can be understood in terms of the analysis presented here.  Other
simulations of the SCDM model (ZANM97) found that the
$b$-distribution matched well, however they only compared the results
for their relatively low-resolution top grid (see
Table~\ref{table:simulations}) which, as we have shown, overestimates
the resulting $b$-parameters.  Similarly, we would argue that the SPH
simulations of this model (\cite{her96}; \cite{dav97}) predicted
$b$-values that were somewhat too large (a conjecture confirmed by
\cite{the98}).  Finally, turning to the last two entries in
Table~\ref{table:simulations}, the work of Miralda-Escud\'e \etal
(1997)\markcite{mcor} had sufficient resolution to produce accurate
$b$-values, and indeed their Figure~17b shows a $b$-distribution which
certainly appears to have a median below that observed.  However, the
cosmological model is substantially different so a direct comparison
is difficult.

How can we restore agreement?  

One possible answer is to find a way to increase the temperature of
the gas.  While this will help by adding to the thermal broadening, it
is clear that this will not be sufficient for the large $b$ lines.
Another way to see this is, if thermal broadening dominated, then our
calculations would predict a correlation between $b$ and $N_{HI}$
since there is a correlation between $T$ and $N_{HI}$; none is
seen. Instead, the increased thermal pressure must act to widen the
physical structures which give rise to the lines (this also means that
we cannot mimic the effect in the simulations by simply increasing the
temperature at the analysis stage).  Figure~\ref{fig:mean_line1} shows
that such a process will require high temperatures, particularly for
the larger column density lines.

Such a large increase (at least a factor of two) in the temperature
seems to be difficult to accomplish.  One way is to ionize He around
$z \sim 3$, but this only appears to boost the temperature by at most
20\% (\cite{hui97b}; \cite{hae97}).  Delaying reionization also helps
because the gas retains some memory of its initial temperature, but
even runs with sudden reionization at $z = 5$ don't change the $b$
distribution significantly.  Similarly, increasing $\Omega_b$ boosts
the equilibrium temperature, but because $T \sim \Omega_b^{1/1.7}$,
even a factor of two change in $\Omega_b$ --- which conflicts with
big-bang nucleosynthesis --- only increases the temperature by 50\%,
which is probably not enough.  All runs in this paper have included a
non-equilibrium chemical model, which increases the temperature
slightly over the assumption of equilibrium ionization, but clearly
not enough.  We have, however, assumed that the radiation field is
uniform.  This approximation will modify the thermal history 
somewhat, but seems unlikely to have a large effect (\cite{mei94};
\cite{mir94}; \cite{gir96}).

One last way to increase the gas temperature is to use thermal energy
from star formation which, of course, we do not include in the
simulations.  This is difficult to rule out, but would require a
remarkably high stellar energy input and the low metallicity of the
low column density gas seems to argue against it (\cite{dav98};
\cite{gir96}).

Another way is to change the density distribution directly, through
the power spectrum or cosmology (\cite{hui98}).  We have only examined
the standard cold dark matter model, and it is possible that other
models will be more successful.  We will address this topic in a
future paper.




\vspace{0.2cm} 

This work is done under the auspices of the Grand Challenge Cosmology
Consortium and supported in part by NSF grants ASC-9318185 and NASA
Astrophysics Theory Program grant NAG5-3923.  We thank Tom Abel for
the use of his rate coefficients, as well as Avery Meiksin and
Jeremiah Ostriker for helpful comments.  MLN acknowledges useful
conversations with Martin Haehnelt.  We thank the referee, David
Weinberg, for many helpful suggestions that improved the quality of
this work.

\end{document}